\documentclass[journal]{IEEEtran}
\usepackage{times}
\usepackage{epsfig}
\usepackage{amsmath}
\usepackage{amsfonts}
\usepackage{graphicx}
\usepackage{amssymb}
\usepackage{amstext}
\usepackage{latexsym}
\usepackage{color}
\usepackage{ifthen}
\usepackage{multirow}
\usepackage{verbatim}
\usepackage{array,tabularx}

\newcommand{\be}{\begin{equation}}
\newcommand{\ee}{\end{equation}}
\newcommand{\bear}{\begin{eqnarray}}
\newcommand{\eear}{\end{eqnarray}}
\newcommand{\bears}{\begin{eqnarray*}}
\newcommand{\eears}{\end{eqnarray*}}
\newcommand{\bi}{\begin{itemize}}
\newcommand{\ei}{\end{itemize}}
\newcommand{\ben}{\begin{enumerate}}
\newcommand{\een}{\end{enumerate}}

\newcommand{\black}{\color{black}}

\newtheorem{theorem}{Theorem}

\newtheorem{proposition}[theorem]{Proposition}
\newtheorem{definition}[theorem]{Definition}
\newtheorem{remark}[theorem]{Remark}

\newtheorem{ex}[theorem]{Example}

\newcommand{\mbf}{\mathbf}

\begin{document}

\title{Securing Dynamic Distributed Storage Systems against Eavesdropping and Adversarial Attacks}

\author{Sameer~Pawar, Salim~El~Rouayheb,~\IEEEmembership{Member,~IEEE}
        and~Kannan~Ramchandran,~\IEEEmembership{Fellow,~IEEE}

\thanks{This research was funded by an NSF grant
(CCF-0964018), a DTRA grant (HDTRA1-09-1-0032), and in part by an
AFOSR grant (FA9550-09-1-0120).}
\thanks{Sameer Pawar is with the Wireless Foundation, Department of Electrical
Engineering and Computer Science, University of California,
Berkeley, CA 94704 USA (e-mail: spawar@eecs.berkeley.edu).}

\thanks{Salim El Rouayheb is with the Wireless Foundation, Department of Electrical
Engineering and Computer Science, University of California,
Berkeley, CA 94704 USA (e-mail: salim@eecs.berkeley.edu).}

\thanks{K. Ramchandran is with the Wireless Foundation, Department of Electrical
Engineering and Computer Science, University of California,
Berkeley, CA 94704 USA (e-mail: kannanr@eecs.berkeley.edu).}

}

\maketitle




\begin{abstract}
We address the problem of securing distributed storage systems against eavesdropping and adversarial attacks. An important   aspect of these systems is node failures over time, necessitating, thus, a repair mechanism in order to maintain a desired high system reliability.   In such dynamic settings, an important security problem is to safeguard the system  from  an intruder who may come at different time instances during  the lifetime of the  storage system  to observe and possibly alter the data stored on some nodes. In this scenario, we give upper bounds on   the maximum amount of information that can be stored safely on the system. For an important operating regime of the distributed storage system, which we call the \emph{bandwidth-limited regime}, we show that our upper bounds are tight and provide explicit code constructions. Moreover, we provide a way to short list the malicious nodes and expurgate the system.
\end{abstract}





\section{Introduction}\label{sec:Intro}
{\em Distributed storage systems} (DSS) consist of a collection of
$n$ data storage nodes, typically individually unreliable, that are
collectively used to reliably store data files over long periods of
time. Applications of such systems are innumerable and include large
data centers and peer-to-peer file storage systems such as OceanStore
\cite{Ocean}, Total Recall \cite{TotalRecall} and DHash++ \cite{DHT}
that use a large number of nodes spread widely across the Internet.
To satisfy important requirements such as data reliability and load balancing, it is desirable for  the system to be designed to enable  a user,  also referred
to as a  data collector,   to download a file stored on
the DSS by  connecting to a smaller number $k$,  $k<n$,  nodes. An important  design problem for such systems arises  from  the individual unreliability of the system nodes due to
 many reasons, such as   disk failures (often due to the use of inexpensive ``commodity'' hardware) or  peer ``churning'' in peer-to-peer storage systems.
In order to maintain a high  system  reliability, the data is stored
redundantly across the storage nodes. Moreover,   the system   is
repaired every time a node  fails by replacing it with a new
node that connects to $d$ other nodes  and download
data to replace the lost one.
\begin{figure}[t]
\begin{center}
\resizebox{2.5in}{2.3in} {\input{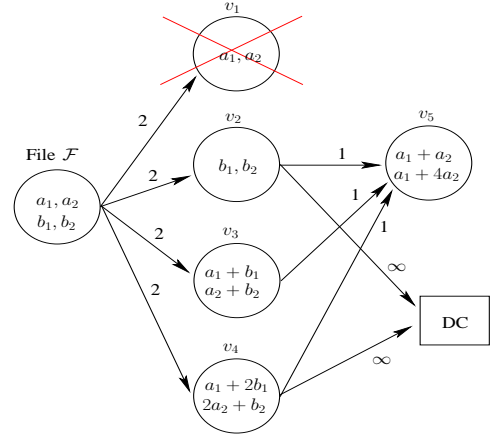}} \caption{ An
example of a distributed data storage system under repair. A file
$\mathcal{F}$ of 4 symbols $(a_1,a_2,b_1,b_2)\in \mathbb{F}_5^4$ is
stored on four nodes using a $(4,2)$ MDS code. Node $v_1$ fails and is
replaced by a new node $v_5$ that downloads $(b_1+b_2),
(a_1+a_2+b_1+b_2)$ and $(a_1+4a_2+2b_1+2b_2)$ from nodes $v_2,v_3$ and
$v_4$ respectively to compute and store $(a_1 + a_2, a_1+4a_2)$.
Nodes $v_2,\dots,v_5$ form a new $(4,2)$ MDS code. The edges in the graph are
labeled by their capacities. The figure also depicts a data collector
connecting to nodes $v_2$ and $v_4$ to recover the stored file.}
\label{fig:distributed_system}
\end{center}
\end{figure}

Codes for protecting data from erasures have been well studied in
classical channel coding theory, and can be used here to increase the
reliability of distributed storage systems.
Fig.~\ref{fig:distributed_system} illustrates an example where a
$(4,2)$ \emph{maximal distance separable} (MDS) code is used to store a file
$\mathcal{F}$ of 4 symbols $(a_1,a_2,b_1,b_2)\in \mathbb{F}_5^4$
distributively on $n=4$ different nodes, $v_1,\dots,v_4$, each having
a storage capacity of two symbols.  The $(4,2)$ MDS code ensures that a data
collector connecting to any $k=2$ storage nodes, out of $n=4$, can reconstruct the
whole file $\mathcal{F}$. However, what distinguishes the scenario
here from the erasure channel counterpart is that, in the event of a
node failure, the system needs to be repaired by replacing the failed
node with a new one. A straightforward repair mechanism would be to
add a replacement node that connects to $k=2$ other nodes, downloads
the whole file, reconstructs the lost part of the data and stores it.
One drawback of this solution is the relatively high repair
bandwidth, {\em i.e.}, the total amount of data downloaded by the new
replacement node.  For this straightforward repair scheme, the repair bandwidth is equal to
the size of the file $\cal{F}$ which can be large in general. A more
efficient repair scheme that requires less repair bandwidth is
depicted in Fig.~\ref{fig:distributed_system} where node $v_1$ fails
and is replaced by node $v_5$. By making node $v_5$ connect to $d=3$
nodes instead of $k=2$, it is possible to decrease the total repair
bandwidth from 4 to 3 symbols. Note that, in the proposed repair
solution, $v_5$ does not store the exact data that was on $v_1$; the
only required property is that the data stored on all the surviving
nodes $v_2,v_3,v_4$ and $v_5$ form a $(4,2)$ MDS code. The above important
observations were the basis of the original work of \cite{DGWWR07}
where the authors showed that there exists a fundamental tradeoff
between the storage capacity at each node and the repair bandwidth.
They also introduced and constructed {\em regenerating codes} as a
new class of codes that generalizes classical erasure codes and
permits the operation of a DSS at any operational point on the
optimal tradeoff curve.

When a distributed data storage system is formed using nodes widely
spread across the Internet, e.g., peer-to-peer systems,
individual nodes may not be secure and  may be thus susceptible to an
intruder that can eavesdrop on the nodes and possibly modify their
data, e.g., viruses, botnet, etc.  In this work, we address the
issue of securing dynamic distributed storage systems,  with nodes continually leaving and
joining the system,  against such intruders.
The dynamic behavior of the system can jeopardize the data by making
the intruder more powerful. For instance, while eavesdropping on a new node during
the repair process, the intruder can observe not only its stored content but
also all its downloaded data. Moreover, it allows an adversary to
introduce errors on nodes beyond his/her control by sending erroneous
messages when contacted for repair.

 In our analysis, we focus on three different types of
intruders: (i) a {\em passive eavesdropper} who can eavesdrop on
$\ell$ nodes in the system, (ii) an {\em active omniscient
adversary} who has complete knowledge of the data stored
in the system and can maliciously modify the data on any $b$
nodes in the system, and (iii) an {\em active limited-knowledge
adversary} who can eavesdrop on any $\ell$ nodes and can maliciously corrupt the data
on any  $b$  nodes among the $\ell$ observed ones. In the last case, the intruder's knowledge about the stored data in the system is limited to what  can be inferred from the nodes he/she is observing.

 We define the {\em secrecy} and {\em resiliency capacities} of a distributed storage system as the maximum
amount of information that it can  store safely,
respectively, in the presence of an eavesdropper or a malicious
adversary. For these intruder scenarios, we derive general upper
bounds on the secrecy and resiliency capacity of the system.
Motivated by system considerations, we define an important
 operation regime that we call the {\em bandwidth-limited} regime
where there is a fixed allowed budget for the repair bandwidth with
no constraints on the node storage capacity. This regime is of
increasing importance  due to the asymmetry in the cost of bandwidth vs. storage. For the
bandwidth-limited regime, we show that our upper bounds are tight
and  provide explicit constructions of  capacity-achieving codes.

The work in this paper is related to the recent work in the literature on secure network coding for networks with restricted wiretapping sets \cite{CHK10} and networks comprising traitor nodes \cite{KTT09}. The problem of studying such networks is known to be much harder in general than  models  considering (unrestricted) compromised edges instead of  nodes. For instance, the work of  \cite{CHK10} implies that finding the secrecy capacity of networks with wiretapped nodes is an NP-hard problem. Moreover,  non-linear coding at intermediate network nodes may be  necessary for securing networks against malicious nodes as shown in  \cite{KTT09}. The contribution of this paper resides, at a high level, in showing that the networks representing distributed storage systems have structural symmetry that makes the security problem more tractable than in general networks. We leverage this fact  to derive the exact expressions of the secrecy and resiliency capacities of these systems in the important bandwidth-limited regime.  Moreover, we present capacity-achieving codes that are linear. These  codes are characterized by a separation property: the file to be stored is first encoded  for security then  stored in the system without any  modification to  the internal operation of the  system nodes.  An additional interesting property of our proposed codes is that, in the active
adversary case, they permit the identification of a small list of suspected nodes guaranteed to contain the malicious ones, permitting  thus the expurgation of the system.


The rest of this paper is organized as follows. In
Section~\ref{sec:related_work}, we discuss related work on
distributed storage systems and secure network coding. In
Section~\ref{sec:model}, we describe the flow graph model for
 distributed storage systems and elaborate on the intruder model. We
provide a brief summary of our main results in
Section~\ref{sec:results}. In Section~\ref{sec:Passive_adversary}, we
derive an upper bound on the secrecy capacity of the system and
provide an achievable scheme for the bandwidth-limited regime.
We provide a similar analysis for the omniscient and limited-knowledge adversary cases respectively in
Section~\ref{sec:omniscient_adversary} and
Section~\ref{sec:limited_adversary}, where we find upper bounds on
the resiliency capacity and construct capacity achieving codes for
the bandwidth-limited regime. We conclude the paper in
Section~\ref{sec:conclusion} and discuss some related open problems.




\section{related Work}\label{sec:related_work}
The pioneering work of Dimakis et al. in \cite{DGWWR07, DGWR07, WDR07}, demonstrated the fundamental trade-off between repair bandwidth and storage cost in a distributed storage system, where nodes fail over time and are repaired to maintain a desired system reliability. They also introduced \emph{regenerating codes} as codes that are more efficient than classical erasure codes for distributed storage applications. In many scenarios of interest, the data is required to exist in the system always in a
systematic form. This has motivated the study of \emph{exact regenerating
codes} \cite{RSKR09, WD09, SR09, SRKR10} that achieve this goal by repairing a failed node with an exact copy of the lost data. The construction of exact regenerating codes in \cite{RSKR09} turns out to be instrumental in achieving the secrecy and resiliency capacity of a DSS in the bandwidth-limited regime.

In \cite{DGWR07}, the construction of regenerating codes was linked
to finding network codes for a suitable network. Network
coding was introduced in the seminal paper of \cite{ACLY00} and
extends the classical routing approach by allowing the intermediate
nodes in the network to encode their incoming packets as opposed to
just copying and forwarding it. The literature on network coding is
now rich in interesting results which can be found in references
\cite{FS07} and \cite{YLC06}, that provide a comprehensive overview
of this area.

In this paper, we are interested in securing distributed storage systems under repair dynamics, which is a special case of the more general problem of achieving security in dynamical systems.
A node-based intruder model is natural in this setting and is related to the recent work of \cite{DDH10} on securing distributed storage systems in the presence of a trusted verifier and that of   Kosut et al. in \cite{KTT09} on protecting data in networks with traitor nodes.
An intruder model that can observe and/or change the data on links, as opposed to nodes, has been extensively studied in the network coding literature. Cai and Yeung introduced in \cite{CY02, CY11} the problem of designing secure network codes in the presence of an eavesdropper, which was further studied
in \cite{FMSS04,RS07,SK08, CHK10}. A
\emph{Byzantine} adversary that can maliciously introduce errors on the network links
was investigated in \cite{HLKMEK04,JL07,JLKHKME08,YSJL10, KHMA10}. The problem of error
correction in networks was also studied by Cai and Yeung in
\cite{YC106,CY206} from a classical coding theory perspective. A different approach for correcting errors in networks
was proposed by Koetter and Kschischang in \cite{KK08},
where communication is established by transmitting subspaces instead
of vectors through the network. The use of maximum rank-metric codes for error control under this model was investigated in \cite{SKK08}.




\section{Model}\label{sec:model}
\subsection{Distributed Storage System}
A distributed storage system (DSS) is a dynamic network of storage
nodes. These nodes include  a source node that has an incompressible
data file $\mathcal{F}$ of $R$ symbols, or units, each belonging to a
finite field $\mathbb{F}$. The source node  is connected to  $n$
storage nodes $v_1,\dots,v_n,$ each having a storage capacity of
$\alpha$ symbols, which may be utilized to save coded parts  of the
file $\mathcal{F}$. The storage nodes are individually unreliable and
may fail over time. To guarantee a certain desired level of
reliability, we assume that the DSS is required to always have $n$
active, {\em i.e.}, non-failed, storage nodes that are simultaneously in service.
Therefore, when a storage node fails, it is   replaced by
a new node with the same storage capacity $\alpha$.   The DSS should be
designed in such a way as to allow any legitimate user or  data collector, that contacts any $k$ out of the $n$ active
storage nodes available at any given time, to be able to reconstruct
the original file $\mathcal{F}$. We term this condition as the
\emph{reconstruction property} of distributed storage systems.

We assume that nodes fail one at a time
\footnote{ Multiple nodes  failing simultaneously  is a rare event. When this occurs, the DSS implements an ``emergency'' recovery process that employs a reserved set of trusted nodes, guaranteed not to be compromised. The trusted nodes then replace the failed ones by acting as data collectors and downloading data from $k$ active nodes. The trusted nodes then consecutively leave the system, thus triggering multiple rounds of the repair process.  },
and we denote by $v_{n+i}$
the new replacement node added to the system to repair the $i$-th
failure. The new replacement node connects then to some $d$
nodes, $d\geq k$, chosen, possibly randomly, out of the remaining
active $n-1$ nodes and downloads $\gamma$ units of data in total from them,
which are then possibly compressed (if $\alpha<\gamma$) and stored on
the node. The data stored on the replacement node can  be
different than the one that was stored on the failed node, as long as
the reconstruction property of the DSS is retained.
The process of replenishing redundancy to maintain the reliability of
a DSS is referred to as the \emph{``regeneration"} or
\emph{``repair"} process, and we call $\gamma$, the total amount of data (in symbols) downloaded for repair,  the  {\em repair bandwidth}
of the system.

 Due to load balancing and ``fairness'' requirements in the system, the repair process is typically  \emph{symmetric}
where the new replacement node downloads equal amount of data,  $\beta=\gamma/d$
units, from each of the node participating in the repair process. We will adopt the symmetric  repair model throughout this paper.
A distributed storage system $\mathcal{D}$ is thus characterized  as
$\mathcal{D}(n,k,d)$, where $k \leq d \leq n-1$. For instance, the DSS depicted in
Fig.~\ref{fig:distributed_system} corresponds to
$\mathcal{D}(4,2,3)$ operating at the point
$(\alpha, \gamma)=(2,3)$.

\subsection{Flow Graph Representation}
\begin{figure}[t]
\begin{center}
\resizebox{3in}{!} {\input{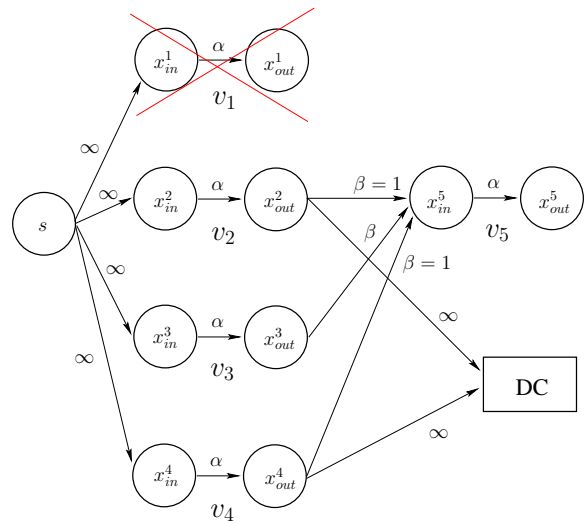}} \caption{The flow
graph model of the DSS ${\cal D}(4,2,3)$ of
Fig.~\ref{fig:distributed_system} when node $v_1$ fails and is replaced
by node $v_5$. Each storage node $v_i$ is represented by two nodes $x_{in}^i$ and $x_{out}^i$ connected by an edge $(x_{in}^i,x_{out}^i)$  of capacity $\alpha$ representing the node storage constraint. A data collector DC connecting to nodes $v_2$ and $v_4$ is also depicted.}\label{fig:FlowGraph}
\end{center}
\end{figure}

We adopt the same model as in \cite{DGWWR07} where  the
distributed storage system is represented by an information flow
graph $\mathcal{G}$. The graph $\mathcal{G}$ is a directed acyclic
graph with capacity constrained edges. It consists of three kinds of
nodes: a  single source node $s$, input storage nodes $x^i_{in}$ and
output storage nodes $x^i_{out}$, and data collectors DC$_j$ for $i,j \in \{1,2,\dots\}$. The source node $s$ holds an information source
$S$ having  the file $\mathcal{F}$ as a special realization. Each
storage node $v_i$ in the DSS is represented by two nodes $x^i_{in}$
and $x^i_{out}$ in ${\cal G}$. To account for the storage capacity of
$v_i$, these two nodes are joined by a directed edge $(x^i_{in},x^i_{out})$ of capacity
$\alpha$ (see Fig.~\ref{fig:FlowGraph}).

The repair process that is initiated every time a failure occurs,
causes the DSS, and consequently the flow graph, to be dynamic and
evolving with time. At any given time, each node  in the graph is
either active or inactive depending on whether it has failed or not.
The graph   $\mathcal{G}$ starts with only the source node $s$ and the nodes $x^1_{in},\dots, x^n_{in}$ connected respectively to the nodes $x^1_{out},\dots,x^n_{out}$. Initially, only the source node $s$ is active and is connected to the storage input nodes $x^1_{in},\dots, x^n_{in}$ by outgoing edges of infinite capacity.  From this point
onwards, the  source node $s$ becomes and remains inactive, and the
$n$ input and output storage nodes become active. When a node $v_i$
fails in a DSS, the corresponding nodes $x_{in}^i$ and $x_{out}^i$
become inactive in $\mathcal{G}$. If a replacement node $v_j$ joins
the DSS in the process of repairing  a failure and connects to $d$
active nodes $v_{i_1},\dots,v_{i_d}$, the corresponding nodes
$x_{in}^j$ and $x_{out}^j$  with the edge $(x_{in}^j,x_{out}^j)$ are
added to the flow graph $\mathcal{G}$, and node $x_{in}^j$ is
connected to the nodes $x_{out}^{i_1},\dots,x_{out}^{i_d}$ by
incoming edges of capacity $\beta=\gamma/d$ units each.  A data collector is
represented   by a  node connected to $k$ active storage output nodes
through infinite capacity links enabling it to download all their stored data and reconstruct the file
$\mathcal{F}$. The graph $\mathcal{G}$ constitutes a multicast
network with the data collectors as destinations. An underlying
assumption here is that the flow graph corresponding to a distributed
storage system depends on the sequence of failed nodes. As an
example, we depict in Fig.~\ref{fig:FlowGraph}
 the flow graph corresponding to the DSS $\mathcal{D}(4,2,3)$ of the previous section (see Fig.~\ref{fig:distributed_system}) when node $v_1$ fails.

Let $\cal{V}$ be the set of nodes in the flow graph $\cal{G}$.  A cut $C(V,\overline{V})$ in the flow graph  separating the   source $s$ from  a data collector DC$_i$ is a partition of the node set of $\cal{G}$ into two subsets $V \subset {\cal V}$ and $\overline{V}=\mathcal{V} \setminus V$, such that $s\in V$ and DC$_i\in \overline{V}$. We say that an edge $(n_1,n_2)$ belongs to a cut $C(V,\overline{V})$ if $n_1\in V$ and $n_2\in\overline{V}$. The \emph{value} of a cut is the sum of the capacities of the edges belonging to it.

\subsection{Intruder Model}
We assume the presence of an illegitimate intruder in the DSS who can eavesdrop on some of the storage nodes, and possibly alter the stored data on some of them in order to sabotage the system. We characterize the power of an intruder by two parameters $\ell$ and $b$, where $\ell$ denotes the number of nodes that the  intruder can eavesdrop on, and $b$ denotes the number of nodes it can control by  maliciously corrupting its  data. We distinguish among three categories of intruders: a \emph{passive eavesdropper} ``Eve'',  an  \emph{active omniscient adversary} ``Calvin'', and an \emph{active limited-knowledge adversary} ``Charlie". We always assume that all the data collectors and intruders have the complete knowledge of the storage and the repair scheme implemented in the  system.


\paragraph{Passive Eavesdropper}
We assume that the eavesdropper Eve can access up to $\ell$, $\ell<k$, nodes of her choice among all the storage nodes, $v_1,v_2,\dots,$ possibly at different time instances as the system evolves. Eve is   passive and can only read the data on the observed $\ell$ nodes without modifying it, {\em i.e.}, $b=0$.
In the flow graph model, Eve is an eavesdropper that can access a fixed
number $\ell$ of nodes chosen from the storage input nodes
$x_{in}^1,x_{in}^2,\dots$. Notice that while a data collector
observes the output storage nodes, {\em i.e.}, the data stored on the nodes
it connects to, Eve, has access to the input storage nodes, and thus
can observe, in addition to the stored data, all the messages incoming to
these nodes. As a result, Eve can choose some of the compromised $\ell$ nodes to be among the initial $n$ storage nodes,  and/or, if she deems it more profitable, she can wait for certain failures to occur and then eavesdrop on the replacement nodes by observing its downloaded data.

\paragraph{Active Omniscient Adversary}
The active adversary Calvin is omniscient \cite{JLKHKME08}, {\em i.e.}, he knows the file $\mathcal{F}$ and the data stored on all the nodes. Moreover, Calvin can control $b$ nodes in total, where $2b < k$, that can include some of the original nodes $v_1,\dots, v_n$, and/or some replacement nodes $v_{n+1},\dots$. Calvin can maliciously alter the data stored on the nodes under his control. It can also send erroneous outgoing messages when contacted for repair or reconstruction. In the flow graph, this corresponds to controlling a set of $b$ input nodes $\{x^{i_1}_{in},x^{i_2}_{in},\hdots,x^{i_b}_{in}\}$  and the corresponding output nodes $\{x^{i_1}_{out},x^{i_2}_{out},\hdots,x^{i_b}_{out}\}$.

\paragraph{Active Limited-knowledge Adversary}
The active adversary Charlie is not {\em omniscient} but has {\em limited knowledge} about the data stored in the system. In particular, he has a limited eavesdropping capability $\ell$ not sufficient enough to know all the stored data. In addition, Charlie can control $b$ nodes of his choice and  maliciously corrupt their  data. In distributed storage systems, an intruder controlling a node will also observe its data. Therefore, we assume that $b \leq \ell$, and that these $b$ nodes are a subset of the $\ell$ eavesdropped nodes. In the flow graph, this corresponds to eavesdropping on some $\ell$ input nodes  $\{x_{in}^{i_1},\dots,x_{in}^{i_\ell}\}$ and controlling a subset of size $b$ of these nodes and the corresponding output nodes.
A similar model was studied in  \cite{JL07, JLKHKME08,YSJL10} where the  authors consider a limited-knowledge adversary that can eavesdrop and control \emph{edges} rather than \emph{nodes} in  multicast networks.




\section{Results}\label{sec:results}
The primary goal of this work is to   secure  distributed storage
systems with repair dynamics in the presence of different types of intruders: passive eavesdropper, active omniscient adversary and active limited-knowledge adversary. We address the following issues:
\begin{itemize}
\item In the case of a passive eavesdropper, we study   the {\em secrecy capacity} $C_s$ of the
    DSS, \emph{i.e.}, the maximum
amount of data that can be stored on the DSS and delivered to a legitimate data collector  without revealing any information about the
data to the intruder.
\item In the case of  an active adversary, we study the {\em resiliency capacity} $C_r$ of the DSS, \emph{i.e.},
the maximum amount of data that can be stored on the DSS and reliably
made available to a legitimate data collector.
\end{itemize}

For a DSS with  symmetric repair, we provide
upper bounds on the {\em secrecy} capacity and {\em resiliency}
capacity. These bounds are maximized for the
choice of repair degree $d=n-1$. In this case, we provide explicit coding schemes that can achieve these bounds in  the bandwidth-limited regime.  Our  results are summarized in Table~\ref{tab:results}. We also show that for the  active adversary controlling $b$ nodes, our
capacity achieving schemes  can
identify a  list, of size at most $2b$ nodes,  that is
guaranteed to contain the malicious nodes. Thus, the system  can  be expurgated of these corrupt nodes, and thereby
 its resiliency to active adversaries is rejuvenated.

\begin{table*}[t]
\begin{center}
\begin{tabular}{|c|c|c|}
  \hline
  & & \\
  Adversary Model & Upper bound  & Bandwidth limited regime ($\Gamma$)\\
  & $\gamma = d\beta$ & $d = n-1, d\beta = \Gamma$\\
  \hline
  & & \\
  Passive eavesdropper ($\ell < k$, $b=0$) & $C_s(\alpha,\gamma) \leq \sum_{i=\ell+1}^{k} \min\{(d-i+1)\beta,\alpha\}$ & $C^{BL}_s(\Gamma) = \sum_{i=\ell+1}^{k}(n-i)\beta$ \\
  & & \\
  \hline
  & & \\
  Active omniscient adversary ($\ell=k$, $2b < k$) & $ C_r(\alpha,\gamma) \leq \sum_{i=2b+1}^{k} \min\{(d-i+1)\beta,\alpha\}$ & $C^{BL}_r(\Gamma) = \sum_{i=2b+1}^{k} (n-i)\beta$ \\
  & & \\
  \hline
  & & \\
  Active limited-knowledge adversary($\ell$, $b \leq \ell$) & $ C_r(\alpha,\gamma) \leq \sum_{i=b+1}^{k} \min\{(d-i+1)\beta,\alpha\}$ & $C^{BL}_r(\Gamma) = \sum_{i=b+1}^{k} (n-i)\beta$ \\
  & & \\
  \hline
\end{tabular}
\end{center}
\caption{Summary of our capacity results for a DSS ${\cal D}(n,k,d)$, with
$\alpha$ units of storage capacity at each node and $\gamma = d\beta$
repair bandwidth. An adversary is characterized by two parameters:
$\ell$, the  number of nodes it can eavesdrop on, and $b$, the number of nodes
it can control.  $C_s$ and $C_r$ denote the secrecy capacity and  resiliency capacity,
respectively. $\Gamma$ is the upper limit on the repair bandwidth for
the bandwidth-limited regime. Note that if   the conditions on $\ell,b$ specified in the first column are not satisfied, then  $C_s,C_r$ are equal to zero}\label{tab:results}
\end{table*}


The upper bounds in Table~\ref{tab:results} are based on cut arguments over the information flow graph representing the DSS \cite{DGWWR07}. Note that when there is no intruder, \emph{i.e.}, $\ell=b=0$, all the upper bounds in the second column of the Table~\ref{tab:results} collapse to the DSS capacity $M =\sum_{i=1}^{k} \min\{(d-i+1)\beta,\alpha\}$ which was derived in the original work of \cite{DGWWR07}. The upper bound on the secrecy capacity $C_s$, for the case of a passive eavesdropper can be explained intuitively by recognizing that when the DSS knows the identity of the $\ell $ compromised nodes it can discard them and avoid using them for storage. Hence, in the expression of the upper bound on $C_s$, we see a loss of
$\ell$ terms in the summation as compared to the capacity
with no intruder.

The upper bound on the resiliency capacity $C_r$, for the case of an active omniscient
adversary, is similar to the one derived in \cite{KTT09} and can be
regarded as a network version of   the Singleton bound: a redundancy of $2b$ nodes is
needed in order to correct the adversarial errors on $b$ nodes. Whereas, a feasible strategy for the limited-knowledge adversary is to delete the data stored on the $b$ nodes it controls rendering them useless resulting in the corresponding upper bound.  Rigorous proofs of these results will be  provided in the coming sections.
%

To get more insight into the above results for the bandwidth-limited case,  we consider an  asymptotic regime for the DSS where the number of nodes goes to infinity whereas the parameters $k,\ell$ and $b$ are kept constant.  We compute the ratios $C^{BL}_s/M$ and $C^{BL}_r/M$,  where $M$ is  the capacity of the DSS in the absence of any intruder. This ratio for the secrecy capacity is,
\begin{eqnarray}\label{eq:passive_asymp}
\frac{C^{BL}_s(\Gamma)}{M} = \frac{\sum_{\ell+1}^k (n-i)\beta}{\sum_{1}^k (n-i)\beta} \approx 1 - \frac{\ell}{k},
\end{eqnarray}
as $n \rightarrow \infty$. Similarly, for the resiliency capacities, we have for omniscient adversary,
\begin{eqnarray}\label{eq:omni_asymp}
\frac{C^{BL}_r(\Gamma)}{M} &\approx& 1 - \frac{2b}{k}.
\end{eqnarray}
And for limited-knowledge adversary,
\begin{eqnarray}\label{eq:ltd_asymp}
\frac{C^{BL}_r(\Gamma)}{M} &\approx& 1 - \frac{b}{k}.
\end{eqnarray}
Note that these asymptotic ratios are reminiscent of the capacity of the classical wiretap channel  \cite{OW} in the case of a passive eavesdropper \eqref{eq:passive_asymp}, the Singleton bound \cite{Blahut} in the case of omniscient adversary \eqref{eq:omni_asymp}, and the capacity of  the erasure channel \cite{Cover} for the case of limited-knowledge adversary \eqref{eq:ltd_asymp}.






\section{Passive Eavesdropper}\label{sec:Passive_adversary}
In this section, we consider a distributed storage system
$\mathcal{D}(n,k,d)$ in the presence of  a passive intruder ``Eve". As described in
Section~\ref{sec:model}, Eve can eavesdrop on any $\ell < k$ storage
nodes\footnote{When Eve observes $\ell\geq k$ the secrecy capacity of the system is trivially equal to zero since Eve can implement
the data collector's scheme to recover all the stored data.} of her choice  in order to learn information about the
stored file. However, Eve  cannot modify the data on these nodes.  We assume that Eve  has
complete knowledge of the storage and repair schemes implemented in
the DSS. Next, we define the {\em secrecy capacity} of a DSS as  the maximum amount of data that can be stored on a DSS under a \emph{perfect secrecy} requirement, \emph{i.e.}, without revealing any information about it to the eavesdropper.

\subsection{Secrecy Capacity}
Let $S$ be a random variable uniformly distributed over  $\mathbb{F}_q^R$   representing the
incompressible data file of size $R$ symbols at the source node, which is to be stored on the DSS. Thus, we have $H(S)=
R$ (in base $\log_q$). Let $V_{in}:=\{x_{in}^1,x_{in}^2,\dots\}$ and
$V_{out}:=\{x_{out}^1,x_{out}^2,\dots\}$ be the sets of input and
output storage nodes in the flow graph, respectively. For each storage node $v_i$, let
$D_i$ and $C_i$ be the random variables representing  its downloaded
messages and stored content respectively. Thus, $C_i$ represents
the data observed by a data collector DC  when connecting to
node $v_i$. If $v_i$ is compromised while joining the DSS, Eve will observe all its downloaded data   $D_i$, with $H(D_i)\leq \gamma$, and not only what it stores.

Let $V_{out}^{a}$ be the collection of all subsets of $V_{out}$ of
cardinality $k$ consisting of the nodes that are simultaneously active, \emph{i.e.}, not failed,
 at a certain instant in time. For any subset $B$ of
$V_{out}$, define $C_B:=\{C_i:x_{out}^i\in B\}$. Similarly for any
subset $E$ of $V_{in}$, define $D_E:=\{D_i:x_{in}^i\in E\}$. The
reconstruction property at the data collector can be written as
\begin{eqnarray}\label{eq:reconstruction_condition}
H(S|C_{B}) &=&  0\quad \forall B\in V_{out}^{a},
\end{eqnarray} and the perfect secrecy condition implies
\begin{eqnarray}\label{eq:secrecy_condition}
H(S|D_{E}) &=&  H(S)\quad \forall E\subset V_{in} \text{ and } |E|\leq
\ell.
\end{eqnarray}

Given a DSS $\mathcal{D}(n,k,d)$ with $\ell$ compromised nodes, its
secrecy capacity, denoted by $C_s(\alpha,\gamma$), is then defined to
be the maximum amount of data that can be stored in this system such that the reconstruction property in \eqref{eq:reconstruction_condition} and the perfect secrecy condition in \eqref{eq:secrecy_condition}
are simultaneously satisfied for all possible data collectors and
eavesdroppers, {\em i.e.},

\begin{eqnarray}\label{eq:secrecy_capacity}
C_s (\alpha,\gamma):= \sup_{{\small \left.
                              \begin{array}{cc}
                                H(S|{C}_{B}) =  0 & \forall B\\
                                H(S|{D}_{E}) =  H(S) & \forall E\\
                               \end{array}
                            \right.}} H(S),
                            \end{eqnarray}
where $B\in V_{out}^a$,  $E\subset V_{in}$ and $|E| \leq \ell.$

\subsection{Special Cases}\label{sec:special}
Before we proceed to the general problem of determining the secrecy
capacity of a DSS, we analyze two special cases  that  shed
 light on the general problem.

\subsubsection{Static Systems}
A static version of the problem studied here corresponds to a DSS
with ideal storage nodes that do not fail, and hence there is no need
for  repair in the system. The flow graph of this system
constitutes then a well-known multicast network studied in network
coding theory called the combination network \cite[Chap. 4]{YLC06}. Therefore, the static storage problem
can be regarded as a special case of wiretap networks \cite{CY11,
RS07}, or equivalently, as the erasure-erasure wiretap-II channel
studied in \cite{AM09}. The secrecy capacity for such systems is
equal to $(k-\ell)\alpha$, and can be achieved using either
the nested MDS codes of \cite{AM09} or the coset codes of \cite{RS07, OW}.

Even though the above proposed solution is optimal for the static
case, it can have a very poor security performance when applied
directly to  dynamic storage systems experiencing  failures and repairs. For instance,
consider the straightforward  way of repairing a failed node by downloading
the whole file and regenerating the lost data. In this case, if Eve observes the new replacement
node while it is downloading the whole file, she will be able to
reconstruct the entire original data. Hence, no secrecy scheme will
be able to hide any part of the data from Eve, and the secrecy rate would be zero.

The case of static systems highlights the new dimension that the
repair process brings into the secrecy picture of distributed storage systems. The dynamic nature of the DSS renders it intrinsically different from the static counterpart making the repair process a key factor that should be carefully designed in order not to jeopardize the whole stored data.

\subsubsection{Systems Using Random Network Coding} \label{sec:RNC_ex}
\begin{figure}[t]
\begin{center}
\resizebox{3in}{!} {\input{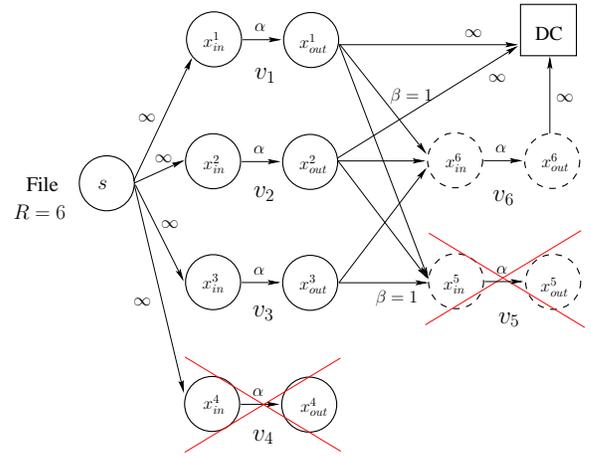}} \caption{The DSS
$\mathcal{D}(4,3,3)$ with $(\alpha,\gamma)=(3,3)$, \emph{i.e.}, $ \beta =1$. Eve
can observe $\ell=2$ nodes. Node $v_4$ fails and is replaced by node
$v_5$, which fails in turn after some time and is replaced by node $v_6$. Nodes
$v_5$ and $v_6$ are compromised and shown with broken boundaries. If random
network coding is used and Eve observes nodes $v_5$ and $v_6$
during repair, it will be able to decode all the stored data with a high probability. }\label{fig:RNC_ex}
\end{center}
\end{figure}

Using the flow graph model, the authors of \cite{DGWWR07} showed that
{\em random linear network codes} over a large finite field  can
achieve any point $(\alpha,\gamma)$ on the optimal storage-repair
bandwidth tradeoff curve with a high probability. Consider an example
of a random linear network code used in a compromised DSS
$\mathcal{D}(4,3,3)$ which stores a file of size $R=6$ symbols with  $\beta = 1$, {\em i.e.}, $\gamma=d\beta=3$, and $\alpha=3$. From \cite{DGWWR07}, it can be shown using the max-flow min-cut theorem  that the maximum file size that can be stored on  this DSS   is equal to $6$ symbols. In this case, each of the initial
nodes $v_1,\dots,v_4$ store $3$ independently generated random linear combinations of the $6$ information symbols. Assume now that node
$v_4$ fails (see Fig.~\ref{fig:RNC_ex}) and is replaced by a new node $v_5$ that connects to $v_1,v_2,v_3$ and downloads from each  $\beta=1$ random linear combination of their stored data. Now
suppose that  node $v_5$  fails after some time and is replaced by
node $v_6$ in a similar fashion. If $\ell = 2$ and Eve had accessed
nodes $v_5$ and $v_6$ while they were being repaired, it would
observe $6$ random linear equations of the data symbols. Since the
underlying field is typically of large size,
the $6$ linear equations observed by  Eve are linearly independent with high probability.
Hence, she will be able to reconstruct the whole file, and  the secrecy rate here is equal to
$0$.  Later in  Example~\ref{ex:DSS_passive_ex} we present a scheme that achieves a secrecy rate of $1$ unit for this DSS.

 While random network codes are  appealing
for use in distributed storage systems due to their decentralized
nature and low complexity, the above analysis shows that this may not
always be the case for achieving security. This is also
in contrast with the case of multicast networks where an intruder can
observe a fixed number of edges instead of nodes \cite{CY11}, wherein,
random network coding performs as good as any deterministic secure
code \cite{SK08}.

\subsection{Results on Passive Eavesdropper} We present here our two main results
for the compromised DSS with {\em passive eavesdropper}:

\begin{theorem}\label{thm:UB_passive} {[Secrecy Capacity Upper Bound]} For a distributed  storage
system $\mathcal{D}(n,k,d)$, with $\ell<k$ compromised nodes, the
secrecy capacity is upper bounded by
\begin{equation}\label{eq:UB}
C_s(\alpha,\gamma) \leq \sum_{i=\ell+1}^{k}
\min\{(d-i+1)\beta,\alpha\},
\end{equation}
\end{theorem}
where $\beta = \gamma/d$.

In the bandwidth-limited regime, we have a constraint on the repair bandwidth $\gamma \leq
\Gamma$, while no constraint is imposed on the node storage capacity
$\alpha$. The secrecy capacity in this regime is thus defined as
\begin{eqnarray}\label{eq:BLregime_UB}
C_s^{BL}(\Gamma)&:=&\sup_{\left.
                                 \begin{array}{c}
                                   \gamma \leq \Gamma\\
                                   \alpha \geq 0\\
                                   \end{array}
                               \right.}C_s(\alpha,\gamma)\\
&\leq& \sup_{\left.
                                 \begin{array}{c}
                                   \gamma \leq \Gamma\\
                                   \end{array}
                               \right.}\sum_{i=\ell+1}^{k}
(d-i+1)\beta.
\end{eqnarray}
The last inequality follows from Theorem~\ref{thm:UB_passive} by setting $\alpha = \Gamma$. When the parameter $d$ is a system design choice, the maximum in the above optimization is attained at $d^*=n-1$. In Section~\ref{sec:passive_achievability}, we demonstrate a scheme that achieves this upper bound, thereby establishing the following theorem.
\begin{theorem}\label{thm:capacity_passive}[Secrecy Capacity: Bandwidth-Limited Regime]
For a distributed data storage system $\mathcal{D}(n,k,d)$ with $d=n-1$ and
$\ell<k$ compromised nodes, the secrecy capacity in the
bandwidth-limited regime is given by
\[
C_s^{BL}(\Gamma) = \sum_{i=\ell+1}^{k} (n-i)\beta,
\]
where $\beta = \frac{\Gamma}{n-1}$ and can be achieved for a node storage capacity  $\alpha = \Gamma$.
\end{theorem}

Before we proceed to prove the above theorems, we consider an example that gives insights into the proof techniques.
\begin{ex}\label{ex:DSS_passive_ex}
Consider again  the DSS $\mathcal{D}(4,3,3)$ operating at $\alpha=3,\beta=1$ and $\ell =2$ of Section~\ref{sec:RNC_ex}. We  show first that the upper bound on the secrecy capacity of this system is $1$ as given by Theorem~\ref{thm:UB_passive}, and then provide a scheme that achieves it.

\begin{figure}[t]
\begin{center}
\resizebox{3in}{!} {\input{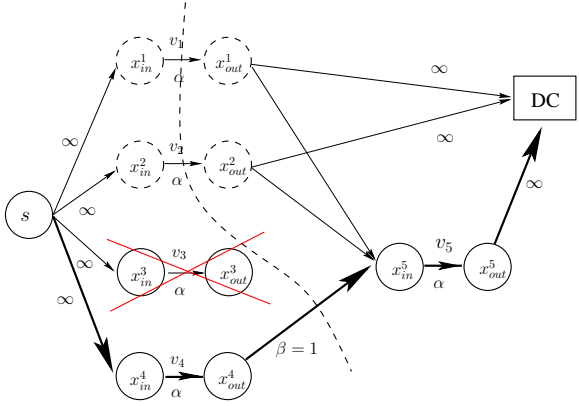}} \caption{The flow
graph of the DSS $\mathcal{D}(4,3,3)$ with $(\alpha,\gamma)=(3,3), \beta=1$ and $\ell=2$. Node $v_3$ fails and is replaced by node $v_5$. Nodes
$v_1,v_2$ are compromised to Eve and are shown with broken
boundaries. A data collector DC connects to nodes $v_1,v_2,v_5$ to
retrieve the data file. The  data collector  can get
at most one unit of information securely on the path
$(s,x^4_{in},x^4_{out},x^5_{in},x^5_{out}, \mbox{DC})$ which is not observed by Eve.}\label{fig:dss_passive_ex}
\end{center}
\end{figure}

To obtain the upper bound on the secrecy capacity, consider the
flow graph of this DSS shown in Fig.~\ref{fig:dss_passive_ex} where
nodes $v_1$ and $v_2$ are compromised and observed by Eve. Suppose that node $v_3$ fails and is replaced by $v_5$ that downloads $\beta=1$ unit of information from each of the $d=3$ nodes $v_1,v_2,v_4$. We focus now on a data
collector that connects to the three nodes $v_1,v_2$ and $v_5$ to reconstruct the source file.  Even if
the source node $s$ and the data collector  knew the location of the eavesdropper, it can
get at most one unit of secure information by ignoring all the
information received from the compromised nodes. The data
 can only be conveyed securely through the path
$(s,x^4_{in},x^4_{out},x^5_{in},x^5_{out}, \mbox{DC})$, that has a ``bottleneck'' edge $(x^4_{out},x^5_{in})$ with capacity $\beta=1$ unit. Since our analysis
is based on a worst case scenario, this gives an upper bound of 1
unit on the secrecy capacity. This bound can be reinterpreted as
taking the minimum value of a cut separating the source $s$ from any data collector in the flow graph after deletion of any two
nodes. This argument can be generalized to any DSS $\mathcal{D}(n,k,d)$ by finding an upper bound on the value of the min-cut in the flow graph after deleting $\ell$ nodes. Thus, we obtain the upper bound of Theorem~\ref{thm:UB_passive} whose  detailed proof is provided in Appendix \ref{app:UB_passive}.
%


Before we  provide a coding scheme that achieves the previous upper bound, we define the \emph{nested MDS codes} \cite{AM09} which will be an important building block in our code construction.

\begin{definition}[Nested MDS Codes]
    An $(n,k)$ MDS code with generator matrix $G$ is called nested if there exists a positive integer $k_0<k$ such that
 $G = \left[
            \begin{array}{c}
              G_1\\
              G_2\\
            \end{array}
          \right]$, with $G_1$, of dimensions $(k_0\times n)$, itself is a generator matrix of an $(n,k_0)$ MDS code.
\end{definition}

 Our proposed capacity-achieving code is depicted in Fig.~\ref{fig:dss_coset_ex} and consists of the concatenation of an outer nested MDS
 code  with a special inner repetition code that was introduced in \cite{RSKR09} for constructing exact regeneration codes.  Let $S\in \mathbb{F}_q$ denote the information symbol that is to be securely stored on the system and ${\cal K} = [K_1 \ \dots \ K_5]$ be a vector of independent random
keys each uniformly distributed over $\mathbb{F}_q$. The MDS coset code is chosen to be a nested MDS code
\cite{AM09} with  its generator matrix given by $G := \left[
        \begin{array}{c}
          G_K\\
          G_S\\
        \end{array}
      \right]$, where
\begin{eqnarray*}
    G_K &=& \left[
      \begin{array}{cccccc}
        1 & 1 & 0 & 0 & 0 & 0\\
        1 & 0 & 1 & 0 & 0 & 0\\
        1 & 0 & 0 & 1 & 0 & 0\\
        1 & 0 & 0 & 0 & 1 & 0\\
        1 & 0 & 0 & 0 & 0 & 1\\
      \end{array}
    \right], \text{ and } \\
G_S &=& \left[ \begin{array}{cccccc}
        1 & 0 & 0 & 0 & 0 & 0
      \end{array}
    \right].
\end{eqnarray*}

 Note that the matrix $G := \left[
        \begin{array}{c}
          G_K\\
          G_S\\
        \end{array}
      \right]$ a generator of a $(6,6)$ MDS code and the sub-matrix $G_K$ is a generator of an $(6,5)$ MDS code ($k_0=5$). Hence, the code generated by $G$ is a nested MDS code. Set, $Z=S + \sum_{i=1}^{5} K_i$, then the codeword $X$  given by

    \begin{eqnarray}\label{eq:coset_code}
    X &=& \left[\begin{array}{cc}
                     {\cal K} & S \\
                   \end{array}
                 \right]\left[
            \begin{array}{c}
              G_K\\
              G_S\\
            \end{array}
          \right],
    \end{eqnarray}
can be written as $X = \left[
 \begin{array}{cccc}
                                                                                         Z & K_1 & \hdots & K_5 \\
                                                                                       \end{array}
                                                                                     \right]$. The encoded symbols $Z,K_1,\dots,K_5$ are then
stored on the nodes $v_1,\dots,v_4$ as shown in Fig.~\ref{fig:dss_coset_ex}, following the special repetition code of Rashmi et al \cite{RSKR09}, which we henceforth refer to as {\em RSKR-repetition code}.

In the RSKR-repetition code used here, nodes $v_1,\dots,v_4$ store respectively $\{Z,K_1,K_2\}$, $\{Z,K_3,K_4\}$, $\{K_1,K_3,K_5\}$ and $\{K_2,K_4,K_5\}$. Since $d=3$, in the case of a failure the new replacement node contacts  all  the $3$ remaining active nodes in the system
and recovers an exact copy of the  lost data. For example,  when node $v_1$ fails  the new replacement node connects to nodes $v_2,v_3$ and $v_4$ and downloads the symbols $Z,K_1$ and $K_2$  from each, respectively. It can also
be checked that a data collector connecting to any $3$ nodes observes
all the symbols $Z,K_1,\dots, K_5$ and hence can decode the information symbol $S$ as $S=Z-\sum_{i=1}^5 K_i$. However, an eavesdropper accessing any two nodes
will observe some subset of $5$ symbols out of $6$, and therefore cannot obtain any information about $S$.
\begin{figure}[h]
\begin{center}
\resizebox{3in}{!} {\input{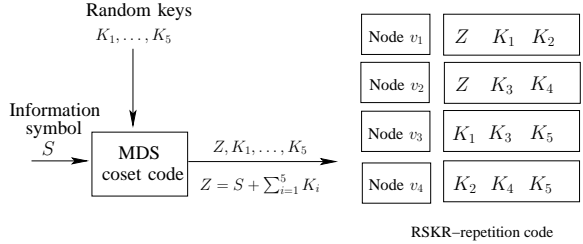}} \caption{ A schematic
representation of the optimal code for the DSS ${\cal D}(4,3,3)$, operating at $(\alpha,\gamma)= (3,3)$ with $\ell =2$, that achieves the secrecy capacity of $1$ unit.
The information symbol $S$ and $5$ independent random keys are mixed
appropriately using an MDS coset code. The encoded symbols are then stored on the DSS using the RSKR-repetition code. An eavesdropper observing any $\ell=2$ nodes cannot get any information about the stored symbol $S$.}\label{fig:dss_coset_ex}
\end{center}
\end{figure}
\end{ex}

In the following section, we provide a generalization
of the code in this example, and show that it achieves the secrecy capacity of DSS for $d=n-1$ in the bandwidth-limited regime,  thus proving Theorem~\ref{thm:capacity_passive}.

\subsection{Secrecy Capacity in the Bandwidth-Limited Regime}\label{sec:passive_achievability}
The special cases studied in Section~\ref{sec:special} pointed out that
the main difficulty in determining the secrecy capacity of
distributed storage systems is due to its dynamic nature. We will demonstrate that in the bandwidth-limited
regime for $d=n-1$, with a careful choice of code, it is
possible to transform the problem of secrecy over a dynamic DSS into
a static problem of secrecy over a point to point channel equivalent
to the erasure-erasure wiretap channel-II in \cite{AM09}. Then, we
show that using nested MDS codes at the source one can achieve the
secrecy capacity of the equivalent wiretap channel.

Our approach builds on the results of \cite{RSKR09} where the authors constructed a family of exact regenerating codes for the DSS $\mathcal{D}(n,k,d)$ with $d=n-1,\alpha = d\beta$. The ``exact" property of these codes allows any repair node to reconstruct and store an identical copy of the data lost upon a failure. The code construction in \cite{RSKR09} consists of the
concatenation of an MDS code with the RSKR-repetition
code. This construction is instrumental for obtaining codes that can achieve the secrecy capacity by carefully choosing the outer  code to be a nested MDS coset code as was done  in Example~\ref{ex:DSS_passive_ex}.

\begin{figure}[t]
\begin{center}
\resizebox{3in}{!} {\input{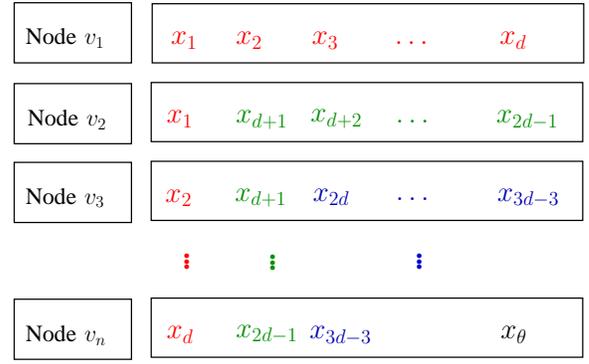}} \caption{ The
structure of the RSKR-repetition code of Rashmi et al
\cite{RSKR09} for $n$ storage nodes, $\alpha = d=n-1,\beta=1$ and $\theta = \frac{n(n-1)}{2}$. The RSKR-repetition code stores $2$ copies of each coded symbol, {\em i.e.,} the total number of stored symbols is $nd = 2\theta$.}\label{fig:Gamma_repetition}
\end{center}
\end{figure}

For simplicity, we will explain the code for $\beta = 1$, {\em i.e.}, $\Gamma=n-1$. For any larger values of $\Gamma$, and in turn of
$\beta$, the file can be split into chunks, each of which can be
separately encoded using the construction corresponding to $\beta =
1$. Since the DSS is operating in the bandwidth-limited regime with no constraint on the node storage capacity, we choose $\alpha = \Gamma$. From \cite{DGWWR07}, we know that for a DSS ${\cal D}(n,k,d=n-1)$ with $\alpha=n-1,\beta=1$ the capacity in the absence of an intruder ($\ell=0$) is $M = \sum_{i=1}^{k}(n-i)$. Let   $R:=\sum_{i=\ell+1}^{k} (n-i)$ be the
maximum number of information  that we could store securely on the DSS, and $\theta:=\frac{n(n-1)}{2}$. Let $S=(s_1,\dots,s_R)\in \mathbb{F}_q^R$ denote the information
file and ${\cal K} = (K_1,\dots,K_{M-R}) \in \mathbb{F}_q^{M-R}$
denote $M-R$ independent random keys each uniformly distributed over
$\mathbb{F}_q$. Then, the proposed code consists of an outer
$(\theta,M)$ nested MDS code (see \eqref{eq:coset_code}) which takes $S$ and ${\cal K}$ as an input and
outputs $X=(x_1,\dots,x_\theta )$, as,
\begin{eqnarray*}
X &=& \left[
        \begin{array}{cc}
          {\cal K} & S
        \end{array}
      \right]\left[
        \begin{array}{c}
          G_K\\
          G_S\\
        \end{array}
      \right],
\end{eqnarray*}
where, $G := \left[
             \begin{array}{c}
               G_K \\
               G_S \\
             \end{array}
           \right]$ is a generator matrix of a $(\theta,M)$ MDS code such that $G_K$ itself is a generator matrix of a $(\theta,M-R)$ MDS code.
This outer $(\theta,M)$ nested MDS code is then followed by an inner
RSKR-repetition code which stores the codeword $X$ on the DSS following the pattern  depicted in Fig.~\ref{fig:Gamma_repetition}.
%

 The RSKR-repetition codes were introduced in \cite{RSKR09} as a method for constructing exact regenerating codes for a distributed storage system. These codes consist of ``filling" the storage nodes
$v_1,\dots,v_n$ successively, by repeating ``vertically" (\emph{i.e},
across all the nodes) the data stored ``horizontally'' ({\em i.e.}, on a single storage node), as shown in  Fig.~\ref{fig:Gamma_repetition}. This procedure can be  described using an auxiliary complete graph over $n$ vertices $u_1,\dots,u_n$
that consists of $\theta $ edges. Suppose the edges are indexed by
the coded symbols $x_1,\dots,x_\theta$. The code then consists of
storing on node $v_i$ the indices of the edges adjacent to vertex
$u_i$ in the complete graph. As a result, the RSKR-repetition code
has a special property that every coded symbol $x_i$ is stored on
exactly two storage nodes, and any pair of two storage nodes have
exactly one coded symbol in common. This property along with the fact that the repair degree $d=n-1$, enables the exact repair of any failed node in the DSS as it was explained in  Example~\ref{ex:DSS_passive_ex}.

The use of the RSKR-repetition code transforms the dynamic
storage system into a static point-to-point channel as explained
below. Notice first that since $\Gamma=\alpha=n-1$,  all the data downloaded during the repair process is stored on the new replacement node  without any further compression\footnote{This corresponds to the \emph{Minimum Bandwidth Regenerating} (MBR) codes described in \cite{DGWWR07}.}. Thus,
accessing a node during repair, {\em i.e.}, observing its
downloaded data, is equivalent to accessing it after  repair, {\em i.e.}, only observing its stored data. Second, the RSKR-repetition code  restore the replacement node with an exact copy of the lost data. Therefore, even though there are failures and repairs, the data
storage system looks exactly the same at any point of time: any data
collector downloads $M$ symbols out of $x_1,\dots,x_\theta$ by
contacting $k$ nodes, and any eavesdropper can observe
$\mu=\sum_{i=1}^{\ell} (d-i+1) = M - R$ symbols. Thus, the system
becomes similar to the erasure-erasure wiretap channel-II of
parameters $(\theta,M,\mu)$\footnote{In the  erasure-erasure wiretap channel-II of
parameters $(\theta,M,\mu)$, the transmitter sends $\theta$ symbols through an erasure channel to a legitimate receiver that receives $M$ symbols. The eavesdropper can observe any $\mu$ symbols out of the transmitted $M$ \cite{AM09}.}. Therefore, since the outer code is a nested
MDS code, from \cite{AM09} we know that it can achieve the secrecy
capacity  of the erasure-erasure wiretap channel which is equal to  $M-\mu$. Hence for the DSS, our codes achieve the secrecy rate of
\[
M - (M-R) = R = \sum_{i=\ell+1}^{k} (n-i).
\]
This rate corresponds to  $\beta=1$. For the general case when $\beta = \Gamma/(n-1)$, the total
secrecy rate achieved  is,
\[
\sum_{i=\ell+1}^{k} (n-i)\beta,
\]
thus completing the proof of Theorem~\ref{thm:capacity_passive}.




\section{Active Omniscient Adversary}\label{sec:omniscient_adversary}
In this section we study distributed storage systems in the presence of an
active  adversary ``Calvin" that can control up to $b$ nodes. Calvin can choose
to control any $b$ nodes among all the storage nodes,
$v_1,v_2,\dots$, and possibly at different time instances as the
system evolves in time due to failures and repairs. Moreover, Calvin is assumed to be
omniscient ($l=k$), so he knows the source file $\mathcal{F}$.  Moreover, since he has
complete knowledge of the storage and repair schemes, he knows the content stored on each node in the system. Under this setting, we define the {\em resiliency capacity}  of a DSS as
the maximum amount of data that can be stored on the DSS and delivered
reliably to any data collector that contacts  any $k$ nodes in the
system.

\begin{figure}[t]
\begin{center}
\resizebox{3.5in}{!} {\input{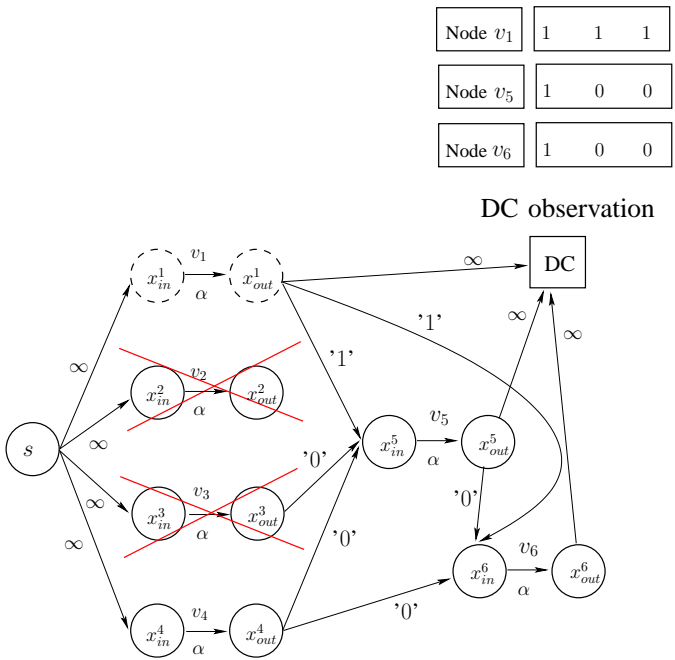}} \caption{Node $v_1$
with broken boundary is compromised and controlled by an omniscient adversary Calvin. Nodes
$v_2$ and $v_3$ fail, and are replaced by nodes $v_5,v_6$ respectively. The all-zero codeword corresponding to message $m=0$ is stored on the DSS. The  Data
collector DC connecting to nodes $v_1,v_5$ and $v_6$ observes a total number  of
$9$ bits out of which $5$ bits are erroneous and equal to ``1'' as shown in the table above.}\label{fig:omni_ex}
\end{center}
\end{figure}

\begin{ex}\label{ex:omni_ex}
Consider again our example of the DSS $\mathcal{D}(4,3,3)$
with $\alpha=\gamma=3$. Assume that there is an
omniscient active adversary Calvin that can control one storage node,
{\em i.e.}, $b=1$, and can modify its stored data  and/or its  messages outgoing to data collectors and
repair nodes.

A first approach for finding a scheme to reliably store data  on this
DSS  would be to use the results in the network coding literature
\cite{JLKHKME08,YC106,CY206,KK08} on the capacity of multicast
networks in the presence of an adversary that can control $t$  edges of unit capacity each. It is shown there that the
resiliency capacity of these networks is equal to $\Omega-2t$, where $\Omega$ is the capacity of the multicast network in the absence of the adversary. This resiliency capacity can be
achieved by overlaying an error-correction code such as a Maximum
Rank Distance  (MRD) code \cite{SK08} on top of the network at the source. This approach turns out to be not very useful here. In fact,  the  capacity in the absence of Calvin is $6$ (see \cite{DGWWR07}),
and $b=1$ corresponds to $t= \alpha = 3$.  Hence, the above approach will
achieve a storage rate of $6-2t = 0$.

We now give a coding scheme that can reliably store 1 bit of information for the DSS. Later, we show that this is also the best that can be
done, {\em i.e.}, the resiliency capacity of this DSS is equal to 1 unit. The proposed code is formed by concatenating a $(6,1)$ repetition code
with an  RSKR-repetition code as shown  in Fig~\ref{fig:omni_ex_code}. The repair process is that of the RSKR-repetition codes described in Section~\ref{sec:passive_achievability}. When a node fails, the replacement node recovers the lost bits by downloading   the bits with same indices from the remaining three active nodes.

\begin{figure}[t]
\begin{center}
\resizebox{3in}{!} {\input{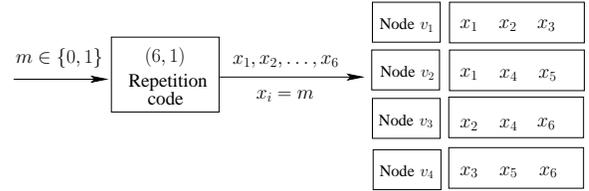}} \caption{
A coding scheme for storing $1$ bit reliably on the DSS ${\cal D}$(4,3,3)
with $\alpha= 3$ bits and $\beta = 1$, in the presence of an omniscient adversary Calvin
who controls $b=1$ node.}\label{fig:omni_ex_code}
\end{center}
\end{figure}

Any data collector contacting three nodes will  observe 9
bits. In the static case, when  no failure or repair occur, only 3 bits
(the ones stored on the compromised node) among the 9 bits observed
by the data collector may be erroneous. In that case,   the DC can
perform a  majority decoding to recover the information bit. However,
in the dynamic model, the DC can receive up to 5 erroneous bits. To
show how this may occur, assume that the DSS is storing the
all-zero codeword, {\em i.e.}, $x_i$ $= 0$ for $i=1,\hdots,6$, in Fig.~\ref{fig:omni_ex_code}, corresponding to the message $m = 0$. Suppose that node $v_1$ is the one that is compromised and  controlled by the adversary Calvin as shown in
Fig.~\ref{fig:omni_ex}. Assume that  Calvin changes all the 3 stored bits $(x_1,x_2,x_3)$ on node $v_1$, from $(0,0,0)$ to $(1,1,1)$ and also sends the erroneous bit ``1'' whenever $v_1$ is contacted for repair.
Now suppose that  node $v_2$ fails and it is replaced by node
 $v_5$ which, based on  the RSKR-repetition structure,   downloads bits $x_1=1, x_4=0$ and $x_5=0$ from  nodes
$v_1,v_3$ and $v_4$ respectively. Suppose also that, after some period of time,
node $v_3$ fails and is replaced by   node $v_6$ which  downloads bits $x_2=1, x_4=0$ and $x_6=0$ from nodes
$v_1,v_4$ and $v_5$ respectively.  An important point to note here is that our repair scheme is fixed and is based on the RSKR-repetition structure irrespective of the possible errors in the bits downloaded during the repair process. As a result a data collector that contacts nodes
$v_1,v_5$ and $v_6$ observes the data as shown in the table in
Fig.~\ref{fig:omni_ex} which includes $5$ errors.

  In a worst case scenario, Calvin will be able to corrupt all the bits in the DSS having the same indices as the bits stored on the  nodes it controls (here the bits with labels $x_1, x_2 $ and $x_3$).  Therefore, Calvin can introduce at most 5 erroneous bits on a collection of  $k=3$ nodes which may be observed by a data collector. In this case, a
majority decoder, or equivalently a minimum Hamming distance decoder, will not be able to decode to the correct message.

To overcome this problem,  we exploit the fact that
Calvin controls only one node, so he can introduce errors only in
specific patterns, to design a special decoder that will always
decode to the correct message $m$ irrespective of Calvin's adversarial
strategy. In fact, for any possible choice of the compromised node, one of the following
four sets $T_1=\{x_4,x_5,x_6\},
T_2=\{x_2,x_3,x_6\},T_3=\{x_1,x_3,x_5\}$ and $T_4=\{x_1,x_2,x_4\}$ is a
\emph{trusted set} that only contains  symbols that were not
altered by Calvin. For example,  when Calvin controls $v_1$, the
trusted set is  $T_1$. The proposed decoder operates in the following way. First, it  finds a  set
$T^*\in\{T_1,\dots,T_4\}$ whose  elements all  agree to either  0 or 1. Then, it
 declares accordingly that message $m=0$ or $m=1$ was stored. This
decoder will always decode to the correct message since each set
$T_i$ intersects with every other set $T_j, j\neq i$, in exactly one
symbol and one of them is a trusted set. Therefore each set
$T_i$ contains at least one symbol which is unaltered by  Calvin.
Thus, if all the symbols in $T_i$ agree, they will agree to
the correct message.
\end{ex}

\subsection{Results on Omniscient Adversary}
In \cite{KTT09}, the resiliency capacity of unicast networks with a
single compromised node was analyzed and a cut-set upper bound was
derived. In the following, Theorem~\ref{thm:cutset} generalizes the bound in \cite{KTT09} for the case of distributed storage systems, where $b\geq 1$ nodes are controlled by an omniscient adversary.

\begin{theorem}\label{thm:cutset} {[Resiliency Capacity Upper Bound]} Consider a distributed storage system DSS
$\mathcal{D}(n,k,d)$. If an omniscient adversary controls any $b\geq 1$ nodes, with $2b < k$,
the resiliency capacity $C_r(\alpha,\gamma)$ is upper bounded as,
\begin{eqnarray}\label{eq:cub}
C_r(\alpha,\gamma) & \leq & \sum_{i=2b+1}^{k}
\min\{(d-i+1)\beta,\alpha\},
\end{eqnarray}
where $\beta = \gamma/d$. If $2b\geq k$, then $C_r(\alpha,\gamma)=0$.
\end{theorem}

This bound is a network version of the Singleton bound and is
obtained by computing the value of certain cuts in the flow graph of
the DSS after the deletion of $2b$ nodes. The detailed proof of the
above theorem is given in Appendix~\ref{app:omni_UB}.

The resiliency capacity in the bandwidth-limited regime is defined as
\[C_r^{BL}(\Gamma):=\sup_{\left.
                                 \begin{array}{c}
                                   \gamma \leq \Gamma\\
                                   \alpha \geq 0\\
                                   \end{array}
                               \right.}C_r(\alpha,\gamma),\]
where $\Gamma$ is the upper limit on the total repair bandwidth.
We again note that if the parameter $d$ is a system design choice, the upper bound of Eq.~\eqref{eq:cub} in the bandwidth-limited regime is maximized for $d=n-1$. In the following section we exhibit a scheme that achieves this upper bound. This result is summarized in Theorem~\ref{thm:omni_capacity}.

\begin{theorem} \label{thm:omni_capacity}
Consider a distributed storage system  $\mathcal{D}(n,k,d=n-1)$
operating in the bandwidth-limited regime. If an omniscient adversary controls $b$ nodes, with $2b < k$, the resiliency capacity of the DSS is given by
\begin{eqnarray}\label{eq:omni_capacity}
C_r^{BL}(\Gamma) & = & \sum_{i=2b+1}^{k} (n-i)\beta,
\end{eqnarray}
where $\beta = \frac{\Gamma}{n-1}$ and can be achieved for a  node storage capacity  $\alpha = \Gamma$. If $2b\geq k$, then $C_r^{BL}(\Gamma)=0.$
\end{theorem}

\subsection{Resiliency Capacity in the Bandwidth-Limited Regime}\label{sec:Omni_capacity}
Similar to the proof of  Theorem~\ref{thm:capacity_passive},
it suffices to show the achievability for $\beta=1$, {\em i.e.},  $\Gamma =
n-1$. In this case, our capacity achieving code uses a node storage
capacity $\alpha =n-1$ symbols.

The code has a similar structure to the scheme used in Section~\ref{sec:Passive_adversary} for the case of a passive adversary and is a generalization of the code used in Example~\ref{ex:omni_ex}. The $(6,1)$ repetition code
in the example is replaced by an $(\theta,R)$ MDS code where $R:=C_r(n-1)=\sum_{i=2b+1}^{k} (n-i)$ and
$\theta=\frac{n(n-1)}{2}$. In the second layer, the output of the MDS code is stored on the DSS following the RSKR-repetition structure as in Fig~\ref{fig:Gamma_repetition}. As explained in Example~\ref{ex:omni_ex}, node failures are repaired using the RSKR-repetition structure (also see Section~\ref{sec:Passive_adversary} for additional details) irrespective of the possible errors introduced by Calvin. Notice that the MDS code used here
has a rate lower then the one used in the passive adversary case in
Section~\ref{sec:passive_achievability} to allow for correcting the
errors introduced by the adversary.

A data collector accessing any $k$ nodes will observe a total of
$\alpha k=(n-1)k$ symbols, out of which $M = \sum_{i=1}^k (n-i)$ symbols have
distinct indices, and $\frac{k(k-1)}{2}$ symbols are repeated due to the
RSKR-repetition code. The adversary can corrupt identically the two copies of each symbol stored on the $b$ controlled nodes. Therefore, the data collector focuses on $M$ symbols with
distinct indices out of $(n-1)k$ and uses them for decoding. These $M$ symbols with distinct indices form a codeword of an $(M,R)$ MDS code, say ${\cal X}$, which are possibly corrupted by the errors introduced by the adversary. The minimum distance
of the  MDS code ${\cal X}$ is,
\begin{equation}\label{eq:dmin_X}
d_{\min}({\cal X}) = M - R + 1 = \sum_{i=1}^{2b} (n-i) + 1.
\end{equation}
The adversary that controls $b$ nodes can introduce up to $t =
\sum_{i=1}^{b} (n-i)$ errors in the set of $M$ symbols with distinct
indices. A simple manipulation shows that $t > \lfloor
\frac{d_{\min}({\cal X}) - 1}{2} \rfloor$.  Therefore, a classical
minimum distance decoder for ${\cal X}$ will not be able to recover
the original file. Thus, the minimum distance decoder fails for this specific adversarial strategy where Calvin corrupts the repeated symbols identically and  cannot be used for a general adversarial strategy.

Next, we present a novel decoder that can correct errors beyond the
classical upper bound of $\lfloor \frac{d_{\min}({\cal X}) -
1}{2}\rfloor$ in the DSS. The main idea is to  take advantage of the special structure of the
error patterns that can be introduced by the adversary.

First, we introduce two definitions that will be useful in describing
the decoding algorithm and that will serve as a generalization of the
concept of  trusted set in the previous example.

\begin{definition} {\bf Puncturing a vector:} Consider a vector $\vec{v} \in \mathbb{F}^{N}$
for some field $\mathbb{F}$. Let $I \subset \{1,2,\hdots,N\}, |I| =
p$, be a given  set. Then {\em puncturing} vector $\vec{v}$ with
pattern $I$ corresponds to deleting the entries in $\vec{v}$
indexed by the elements in  $I$ to obtain a vector $\vec{v}_I \in
\mathbb{F}^{N-p}$.
\end{definition}

\begin{definition} {\bf Puncturing a Code:} Consider a code ${\cal C}$ in  $\mathbb{F}^{
N}$. Let $I \subset \{1,2,\hdots,N\}, |I| = p$, be a given  set.
The {\em punctured code} ${\cal C}_{I}$ is obtained by {\em
puncturing} all the codewords of ${\cal C}$ with pattern $I$, {\em i.e.},
\[
{\cal C}_I := \{\vec{x}_I| \vec{x} \in {\cal C}\}.
\]
\end{definition}

\begin{proposition}\label{prop:punc_MDS} If ${\cal C}$ is an MDS code with parameters $(n,k)$
then for any given fixed pattern $I \subset \{1,2,\hdots,n\}, |I| = p
< (n-k+1)$, the punctured code ${\cal C}_I$ is also an MDS code with
parameters $(n-p,k)$.
\end{proposition}

\subsubsection*{Decoding Algorithm}\label{sec:omni_decoding}
Let $B, |B|\leq b$, denote the set of storage nodes  controlled by the
adversary. Because of the exact repair property of the
RSKR-repetition codes, it is sufficient to focus on the case
when $B\subset \{v_1,\dots,v_n\}$  with $|B|=b$. For each such set
$B$, we define $I_B\subset \{1,2,\hdots,\theta\}$ to be the set of
the indices of the symbols stored on the nodes in $B$. For instance, in  Example~\ref{ex:omni_ex}, if
$B=\{v_1\}$, $I_B=\{1,2,3\}$.

The decoding algorithm proceeds in the following way:
\begin{enumerate}
    \item The data collector connecting to $k$ nodes selects any $M$ symbols with distinct indices,
    out of the $(n-1)k$ observed symbols, as its input $Y \in \mathbb{F}_q^M$ for
    decoding. In Example~\ref{ex:omni_ex}, Fig.~\ref{fig:omni_ex}, the DC connecting to nodes $v_1,v_5,v_6$ observes
    vector $(y_1,y_2,y_3,y_1,y_4,y_5,y_2,y_4,y_6)$. After  removing the repeated symbols,
    we get $Y=(y_1,y_2,y_3,y_4,y_5,y_6)$. Note for a fixed DC, $Y$ is a
    codeword of an $(M,R)$ MDS code which we call ${\cal X}$. $Y$ includes  possible errors introduced by the
    adversary. The code ${\cal X}$ itself is a punctured code of the outer $(\theta,R)$ MDS
    code.

    \item For each $B\subset \{v_1,\dots,v_n\},  |B|=b$, find $I_B$.
    \item Puncture $Y$ and the code ${\cal X}$ with pattern $I_B$ to
    obtain the observed word $Y_{I_B}$ and punctured code ${\cal X}_{I_B}$.
    Note that due to the RSKR-repetition structure, the size of such puncturing pattern is
    \[
    |I_B| = \sum_{i=1}^b (n-i)
    \]
    which is less than the minimum distance of the MDS code ${\cal X}$
    (see \eqref{eq:dmin_X}). Hence, by Proposition~\ref{prop:punc_MDS} ${\cal
    X}_{I_B}$ is an MDS code.

    \item Let $H_{{\cal X}_{I_B}}$ be the parity check matrix of the
    punctured code ${\cal X}_{I_B}$. Compute the syndrome of the observed
    word $Y_{I_B}$ as
    \[
    \vec{\sigma}_{I_B} = H_{{\cal X}_{I_B}}Y_{I_B}^T.
    \]
    \item If $\vec{\sigma}_{I_B} = 0$, then $Y_{I_B}$ is a codeword of ${\cal X}_{I_B}$. Assume it to be a trusted codeword and decode to message using the code ${\cal X}_{I_B}$.
\end{enumerate}

\subsubsection*{Proof of Correctness}
We now prove the correctness of the above decoding algorithm by
showing that it will always correct the errors introduced by the
adversary and output the correct message. Notice first that  the
syndrome $\vec{\sigma}_{I_B}$  will always be equal to zero whenever
$B=B^*$, the actual set of nodes controlled by the  adversary (which
is not known to the data collector). Therefore, the above decoding
algorithm will  always give  an output. Next, we show that this output
always corresponds to the correct message stored on the DSS. Denote by $X$   the true codeword in ${\cal X}$, that  would have been observed by the DC in the absence of Calvin. Let
 $B^*$ be the set of the $b$ traitor nodes. Then, the proposed decoding algorithm fails iff there exists
some other set $B\neq B^*$, and some other codeword $X'\in {\cal X}$, s.t. $X'\neq X$, for which $Y_{I_B} = X'_{I_B} \in {\cal X}_{I_B}$. This implies
that
\begin{equation}\label{eq:d_x_x'}
X_{I_{B^*} \cup I_B} = X'_{I_B\cup I_{B^*}}.
\end{equation}

But, from the RSKR-repetition code structure we know
\begin{eqnarray}\label{eq:card_I_U_I}
|I_{B^*} \cup I_B | & \leq & \sum_{i=1}^{2b}(n-i).
\end{eqnarray}

Equations~\eqref{eq:d_x_x'} and \eqref{eq:card_I_U_I} imply that
$d_{\min}({\cal X}) \leq \sum_{i=1}^{2b}(n-i)$ which contradicts
equation~\eqref{eq:dmin_X}.\\

\begin{remark}[Decoder complexity] The complexity of the
proposed decoder is exponential in the number $b$ of malicious nodes. Therefore, it is not practical for  systems
 with large  values of $b$. However, this decoder can be regarded as a proof technique for the achievability of the resiliency capacity $C_r^{BL}$ of Theorem~\ref{thm:omni_capacity}.
\end{remark}

\begin{remark}\label{rm:expurgation}[Expurgation of malicious nodes] As shown above, the proposed
decoder always decodes to the correct message, and thus, can identify the indices of any erroneous symbols. The data collector can then report this set of  indices
to a central authority (tracker) in the system. This authority will combine all the sets it receives, and knowing the RSKR-repetition structure (see
Fig.~\ref{fig:Gamma_repetition}), it forms a list of suspected  nodes that will surely include the malicious nodes that are sending  corrupted data to the data
collectors.    Since there are at most $b$ malicious nodes and each  symbol $x_i$ is stored on exactly two nodes,   the size of the list will be at most  $2b$. The
system is then purged by discarding the nodes in this list.

\end{remark}




\section{Active Limited-Knowledge Adversary} \label{sec:limited_adversary}
In this section, we consider the case of a non-omniscient active
adversary with limited eavesdropping and controlling capabilities.
We assume the adversary   can eavesdrop on
$\ell$ nodes and control some subset of $b \leq \ell$ nodes out   of these $\ell$
nodes. The  adversary's knowledge about the  stored file is {\em
limited} to what it can deduce from the observed nodes.
Moreover, we assume that the adversary knows the coding and
decoding strategies at every node in the system.  Clearly when $\ell\geq k$, the adversary becomes omniscient. We
are interested here in the limited-knowledge scenario that does not
degenerate into the omniscient model studied in the previous section. For this case, we demonstrate
that the resiliency capacity of the DSS exceeds that of   the omniscient
case, and can be achieved by storing a small {\em hash} on the nodes
in addition to the data. Our approach is similar to that of
\cite{JL07, JLKHKME08, YSJL10}, where the authors consider a limited-knowledge adversary that can eavesdrop and control \emph{edges}
rather than \emph{nodes} in  multicast networks.


%

\begin{ex}\label{ex:Ltd_example}
Consider a DSS ${\cal D}(5,3,4)$ with $\alpha = \gamma = 4$ with an
adversary Charlie that can eavesdrop on and control one node, {\em i.e.},
$b=\ell=1$. In the omniscient case with $b=1$, the resiliency capacity of this system as given by Theorem~\ref{thm:omni_capacity} is equal to $2$. Here, we show that
the limitation on Charlie's knowledge can be leveraged to increase
the resiliency capacity to $5$.

First, we show that  the resiliency capacity for
this DSS is upper bounded by $5$. To that end, consider the case when node $v_1$ is
observed and controlled by Charlie. Moreover, assume that nodes $v_2$
and $v_3$ fail successively and are replaced by nodes $v_6$ and $v_7$
as shown in Fig.~\ref{fig:LtdCalvin_ex}. Consider now a data collector DC
that connects to nodes $v_1,v_6,v_7$ and wants to reconstruct the
stored file. One possible attack that  Charlie can perform, is to erase all the
data stored on node $v_1$, {\em i.e.}, always  change it to a fixed value irrespective of the stored file. This
renders node $v_1$ useless and the system performs as if node $v_1$
was removed which reduces the value of the cut $C(V,\bar{V})$ (see
Fig.~\ref{fig:LtdCalvin_ex}) between the source $s$ and data collector DC to $5$.

\begin{figure}[t]
\begin{center}
\resizebox{3in}{!} {\input{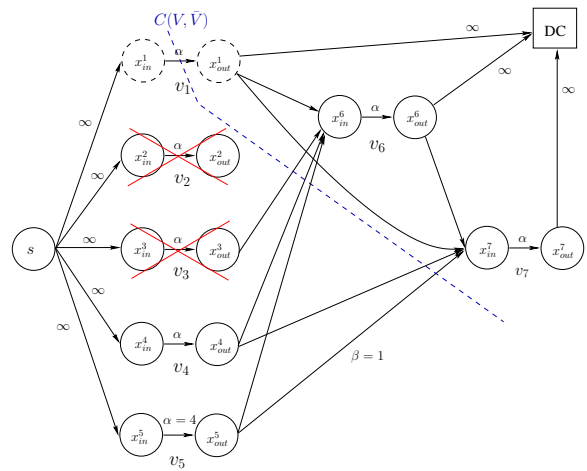}} \caption{The limited-knowledge adversary Charlie
eavesdrops and controls node $v_1$, shown with the broken boundary. If Charlie erases the data stored on node $v_1$, the value of the cut $C(V,\bar{V})$, with $\bar{V} = \{x^1_{out},x^6_{in},x^6_{out}, x^7_{in},x^7_{out},\mbox{DC}\}$, between the source
node $s$ and a data collector DC accessing nodes $v_6,v_7,v_8$ becomes equal to $5$.}\label{fig:LtdCalvin_ex}
\end{center}
\end{figure}

We now exhibit a code that uses   a simple ``correlation'' hash scheme to achieve the above upper bound  with high probability.
\paragraph{Code Construction}
The code consists of an outer $(10,5)$ MDS code over $\mathbb{F}_{q^v}$, followed by the RSKR-repetition code enabling the exact repair of the nodes in the case of
failures. Furthermore, each data packet $\mbf{x_i} \in \mathbb{F}_{q^v}$ is appended with a hash vector $\mbf{h_i}=(h_{i,1},\dots, h_{i,10}) \in \mathbb{F}_q^{10}$
computed as,
\begin{eqnarray*}
{h}_{i,j} &=& \mbf{x_i}\mbf{x_j}^T,
\end{eqnarray*}
for $j = 1,2,\hdots,10$, where with abuse of notation, $\mbf{x_i}$ also denotes the vector $(x_{i,1},\dots,x_{i,v})$ in $\mathbb{F}_q^{v}$ representing the
corresponding element of $\mathbb{F}_{q^v}$. The schematic form of the code is shown in Table~\ref{tab:Ltd_example_code} below.

For simplicity, we assume in this example that the hash values stored on the nodes are made secure from Charlie who can neither observe, nor corrupt them. Later in Appendix~\ref{app:ltd}, we explain how this can be achieved in the general case with a negligible sacrifice in the system capacity. Note
that even though Charlie cannot directly observe the hash table, he
can generate some of the hash values using the observed data packets
on $\ell=1$ eavesdropped nodes, since he knows the coding scheme.  Charlie can use these computed hash values to carefully introduce errors in the data symbols such that it is still consistent with these hash values.

%

\begin{table}[h]
\begin{center}
\begin{tabular}{|c|c|c|}
  \hline
  Node & data $\in \mathbb{F}_{q^v}$ & hash $\in \mathbb{F}_q^{10}$\\
  \hline
  $v_1$ & $\mbf{x_1},\mbf{x_2},\mbf{x_3},\mbf{x_4}$ & $\mbf{h_1},\mbf{h_2},\mbf{h_3},\mbf{h_4}$\\
  \hline
  $v_2$ & $\mbf{x_1},\mbf{x_5},\mbf{x_6},\mbf{x_7}$ & $\mbf{h_1},\mbf{h_5},\mbf{h_6},\mbf{h_7}$\\
  \hline
  $v_3$ & $\mbf{x_2},\mbf{x_5},\mbf{x_8},\mbf{x_9}$ & $\mbf{h_2},\mbf{h_5},\mbf{h_8},\mbf{h_9}$\\
  \hline
  $v_4$     & $\mbf{x_3},\mbf{x_6},\mbf{x_8},\mbf{x_{10}}$ & $\mbf{h_3},\mbf{h_6},\mbf{h_8},\mbf{h_{10}}$\\
  \hline
  $v_5$     & $\mbf{x_4},\mbf{x_7},\mbf{x_9},\mbf{x_{10}}$ & $\mbf{h_4},\mbf{h_7},\mbf{h_9},\mbf{h_{10}}$\\
  \hline
\end{tabular}
\end{center}
\vspace*{0.1in}
\caption{The schematic form of the code stored on the DSS $\mathcal{D}$(5,3,4), along with the secure hash table that is not accessible to the adversary Charlie.}\label{tab:Ltd_example_code}
\end{table}

\paragraph{Decoding logic}
A data collector contacting $3$ nodes observes 12 symbols in total. In a worst case scenario, Charlie can corrupt    6 out of these 12 symbols.  This can happen, for instance, when Charlie eavesdrops and controls node $v_1$, and maliciously changes its stored data from $\mbf{x_i}$   to $\mbf{y_i}=\mbf{x_i}+\mbf{e_i}, \mbf{e_i}\neq 0,  i=1,\dots,4$. Then, $v_2,v_3$ fail successively (as shown in Fig.~\ref{fig:LtdCalvin_ex})  and Charlie sends the erroneous symbols $\mbf{y_1}$ and $\mbf{y_2}$, respectively, to  nodes $v_5$ and $v_6$ during the repair process. In  this scenario,  a data collector, unaware of Charlie's actual node
location, accessing nodes $v_1,v_6$ and $v_7$ will have among its observation $6$ corrupted symbols, namely those having indices $1,\dots,4$ as shown in Table~\ref{tab:DC_observation}, where the symbol $\mbf{y_i}$ denotes the possibly corrupted version of $\mbf{x_i}, i=1,\dots,9$.  Here, we have $\mbf{y_i}=\mbf{x_i}, i=5,\dots,9$. The table also shows the
 hash vectors  observed by the same data collector.

\begin{table}[h]
\begin{center}
\begin{tabular}{|c|c|c|}
  \hline
  Node & data $\in \mathbb{F}_{q^v}$ & hash $\in \mathbb{F}_q^{10}$\\
  \hline
  $v_1$ & $\mbf{y_1},\mbf{y_2},\mbf{y_3},\mbf{y_4}$ & $\mbf{h_1},\mbf{h_2},\mbf{h_3},\mbf{h_4}$\\
  \hline
  $v_6$ & $\mbf{y_1},\mbf{y_5},\mbf{y_6},\mbf{y_7}$ & $\mbf{h_1},\mbf{h_5},\mbf{h_6},\mbf{h_7}$\\
  \hline
  $v_7$ & $\mbf{y_2},\mbf{y_5},\mbf{y_8},\mbf{y_9}$ & $\mbf{h_2},\mbf{h_6},\mbf{h_8},\mbf{h_9}$\\
  \hline
\end{tabular}
\end{center}
\vspace*{0.1in}
\caption{The data symbols and hash values observed by the data collector contacting  nodes $v_1,v_6,v_7$, when node $v_1$ is controlled by Charlie.}\label{tab:DC_observation}
\end{table}

Among the 12 stored symbols $\mbf{x_i}$ observed by the data collector  and their hashes
$\mbf{h_i}$,  each of the 3 symbols with indices  $1,2,5$ and
the corresponding hash vectors $h_1, h_2, h_5$ are repeated twice. Since the adversary can change both copies of each
repeated data symbol identically, our decoder  focuses only on a set of $M = 9$ symbols of distinct indices and the corresponding hash vectors for decoding.  Note that the corresponding $9$ symbols
$(\mbf{x_1}, \hdots, \mbf{x_9})$ form a codeword of a $(9,5)$ MDS code that we refer to as ${\cal X}$.

Let $H$ denote the  $9\times 9$ hash matrix observed  by the data collector, obtained as
\[
H = \left[
      \begin{array}{c}
      \mbf{h_1}\\
      \mbf{h_2}\\
      \vdots\\
      \mbf{h_9}\\
      \end{array}
    \right],
\]
where the $i^{\text th}$ row $\mbf{h_i} \in
\mathbb{F}_q^{10}$ corresponds to the hash vector of the symbol
$\mbf{y_i}, i = 1,\hdots,9$. The data collector then computes its own $9 \times 9$ hash matrix $\hat{H}$ from the $9$ observed symbols
$\mbf{y_i}$ as
\[
\hat{H}_{ij}=\mbf{y_i}\mbf{y_j}^T, \ \ \ \ \ \ 1 \leq i,j\leq 9.
\]
Then, it compares the entries in $\hat{H}$ with the corresponding entries in $H$ to generate a $9 \times 9$ comparison table. Table~\ref{tab:checkmarks} is an example of such a comparison
 table where a ``$\checkmark$" in position $(i,j)$ indicates that the
computed hash and the observed hash match, \emph{i.e.}, $\hat{H}_{ij}=H_{ij}$,
whereas ``$\times$" indicates that $\hat{H}_{ij}\neq H_{ij}$ due to
the errors introduced by the adversary.
\begin{table}[h]
\begin{center}
\resizebox{3.2in}{!}{
\begin{tabular}{|c|c|c|c|c|c|c|c|c|c|}
  \hline
  Data Symbol & $\mbf{y_1}$ & $\mbf{y_2}$ & $\mbf{y_3}$ & $\mbf{y_4}$ & $\mbf{y_5}$ & $\mbf{y_6}$ & $\mbf{y_7}$ & $\mbf{y_8}$ & $\mbf{y_9}$\\
  \hline
  $\mbf{y_1}$ & \checkmark & \checkmark & \checkmark & \checkmark & $\times$ & $\times$ & $\times$ & $\times$ & $\times$ \\
  \hline
  $\mbf{y_2}$ &\checkmark & \checkmark & \checkmark & \checkmark & $\times$ & $\times$ & $\times$ & $\times$ & $\times$ \\
  \hline
  $\mbf{y_3}$ &\checkmark & \checkmark & \checkmark & \checkmark & $\times$ & $\times$ & $\times$ & $\times$ & $\times$ \\
  \hline
  $\mbf{y_4}$ &\checkmark & \checkmark & \checkmark & \checkmark & $\times$ & $\times$ & $\times$ & $\times$ & $\times$ \\
  \hline
  $\mbf{y_5}$  & $\times$ & $\times$ & $\times$ & $\times$ & \checkmark & \checkmark & \checkmark & \checkmark & \checkmark \\
  \hline
  $\mbf{y_6}$  &$\times$ & $\times$ & $\times$ & $\times$ & \checkmark & \checkmark & \checkmark & \checkmark & \checkmark \\
  \hline
  $\mbf{y_7}$  &$\times$ & $\times$ & $\times$ & $\times$ & \checkmark & \checkmark & \checkmark & \checkmark & \checkmark \\
  \hline
  $\mbf{y_8}$  &$\times$ & $\times$ & $\times$ & $\times$ & \checkmark & \checkmark & \checkmark & \checkmark & \checkmark \\
  \hline
  $\mbf{y_9}$  &$\times$ & $\times$ & $\times$ & $\times$ & \checkmark & \checkmark & \checkmark & \checkmark & \checkmark \\
  \hline
\end{tabular}}
\end{center}
\vspace*{0.1in}
\caption{Example of the comparison table of the  hash matrices $H$ and $\hat{H}$.
Note that since Charlie observes the  data symbols
$\{\mbf{x_1},\hdots,\mbf{x_4}\}$, he can introduce errors such that
the hash values of $\{\mbf{y_1}\hdots,\mbf{y_4}$\} are consistent.
}\label{tab:checkmarks}
\end{table}

The decoder selects a \emph{trusted set} of $5$ symbols from
$\{\mbf{y_1},\dots,\mbf{y_9}\}$ that index a
$5\times 5$ sub-table of the comparison table where all the entries
are ``$\checkmark$", \emph{e.g.}, symbols $\mbf{y_5,y_6,y_7,y_8,y_9}$ in Table~\ref{tab:checkmarks}. It then sets the remaining    $4$ symbols as erasures and
proceeds to decode using a minimum distance decoder for the $(9,5)$
MDS code ${\cal X}$, that can correct up to $4$ erasures. There
always exists at least one set of $5$ symbols that generates a
consistent hash table, \emph{e.g.}, $T=\{\mbf{y_5,y_6,y_7,y_8,y_9}\}$ when
Charlie controls  node $v_1$. Hence, the proposed
decoding will eventually stop and output a  decoding decision. Next, we analyze the probability of
selecting a trusted set that results in an error in decoding.

\paragraph{Error Analysis}  Let $E = \{\mbf{x_1},\hdots,\mbf{x_4}\}$ denote the set of data symbols observed by Charlie by eavesdropping on $\ell=1$ node ($v_1$ in this case). The above proposed decoder may result in an error only if the chosen trusted set $T$ contains at least one erroneous symbol, say $\mbf{y_1}$. Therefore, we can write  $\mbf{y_1} =\mbf{x_1} + \mbf{e_1}$ for some error $\mbf{e_1}\neq 0 \in \mathbb{F}_{q^v}$.  Any chosen trusted set $T$ is also guaranteed to contain at least one error-free symbol that is not observed by Charlie, say $\mbf{y_5} = \mbf{x_5} \notin E$. To see this, note that the cardinality of the trusted set $T$ is $5$, and by eavesdropping and controlling any one node Charlie can observe and introduce  errors in a maximum of $4$ symbols with distinct indices to any data collector observation. For the set $T$,  containing $\mbf{y_1},\mbf{y_5}$ along with  $3$ other symbols, to be a trusted set, it has to generate a consistent hash table of size $5 \times 5$. Therefore, Charlie has to pick the error $\mbf{e_1}$  to satisfy $\mbf{x_5}\mbf{e_1}^T = 0$.


The observation $E = \{\mbf{x_1},\hdots,\mbf{x_4}\}$ of Charlie is independent of $\mbf{x_5}$ due to the MDS property of the outer code. Therefore, for any choice
of $\mbf{e_1}$ that Charlie makes, there are $q^{v}$ equally likely choices of $\mbf{x_5}$, out of which $q^{v-1}$ are orthogonal to the chosen $\mbf{e_1}$. Hence,
the consistency condition of hash $\hat{H}_{5,1} = H_{5,1}$ is satisfied with probability,
\[
P_r(\mbf{x_5}\mbf{e_1}^T = 0|E, \mbf{e_1}) = \frac{1}{q}.
\]

Note that if Charlie could observe the complete hash table, then $\mbf{x_5}$ is no more independent of Charlie's observation. For example, if Charlie observes the
hash value $H_{2,5} = \mbf{x_2}\mbf{x_5}^T$, then for a given value of $\mbf{x_2}$ and $H_{2,5}$, there are only $q^{v-1}$ equally likely choices for $\mbf{x_5}$.
In which case Charlie can always choose $\mbf{e_1}$ to belong to the  space orthogonal to   $v-1$ dimensional space of possible choices of $\mbf{x_5}$, thus,
deceiving the proposed decoder. Therefore, it is crucial to keep the hash values secure from Charlie.

It can be verified that the above reasoning easily carries to any
choice of $b=1$ node controlled by Charlie. Therefore,  the
probability of error is upper bounded by $1/q$ which vanishes with
increasing the field size $q$.

\paragraph{Rate Analysis} We encode $5$ information symbols in $\mathbb{F}_{q^v}$ to form the coded symbols $\mbf{x_i}, i = 1,\hdots,10$. For these $10$ symbols
we construct a hash table of size $10 \times 10$ with elements in $\mathbb{F}_q$. Hence the total overhead of the hash table is $\frac{100}{5v} = {\cal
O}(\frac{1}{v})$ per information symbol. Thus, the rate of our code is $5 - {\cal O}(\frac{1}{v})$ which approaches $5$ with an increasing block length $v$.
\end{ex}

\subsection{Results on Active Limited-Knowledge Adversary}
Below we summarize our  two main results on the  resiliency capacity in the case of a limited-knowledge adversary.

\begin{theorem}\label{thm:Ltd_UB} For a DSS ${\cal D}(n,k,d)$ with an adversary that can eavesdrop on any $\ell < k$ nodes and control a subset of size $b$  of these $\ell$ nodes ($b \leq \ell$), the following upper bound holds on the resiliency capacity,
\begin{equation}
C_r(\alpha,\gamma) \leq \sum_{i=b+1}^{k}\min\{(d-i+1)\beta,\alpha\}
\end{equation}
where $d\beta = \gamma$.
\end{theorem}
\begin{proof}(sketch) Consider a case when nodes $v_1,\hdots,v_k$ fail successively and are replaced by nodes $v_{n+1},\hdots,v_{n+k}$ as shown in Fig.~\ref{fig:converse}. Also consider a data collector DC that contacts these $k$ nodes $\{v_{n+1},\hdots,v_{n+k}\}$ to retrieve the source file. If the adversary Charlie controls the $b$ nodes $\{v_{n+1},\hdots,v_{n+b}\}$, one possible adversarial strategy that Charlie can use is to erase all the data stored on these $b$ nodes, {\em i.e.}, always change it to a fixed value irrespective of the file stored on the DSS. This renders the $b$ controlled nodes useless, resulting in the upper bound stated in the theorem.
\end{proof}

Let $R:= \sum_{i=b+1}^{k}\min\{(d-i+1)\beta,\alpha\}$ and ${\cal E}:= \sum_{i=1}^{\ell}\min\{(d-i+1)\beta,\alpha\}$. Our second results  states that if the eavesdropping capability $\ell$ of the adversary Charlie is limited, in particular $\ell$ is such that ${\cal E} < R$, the upper bound in Theorem~\ref{thm:Ltd_UB} can be achieved for $d=n-1$ in the bandwidth-limited regime.

\begin{theorem}\label{thm:LtdCap} Consider a DSS ${\cal D}(n,k,d=n-1)$ operating in the bandwidth-limited regime in the presence of an adversary that can eavesdrop on $\ell$ nodes  and controls a subset of size $b$  of these $\ell$ nodes ($b \leq \ell$). Then, if the adversary is limited-knowledge, {\em i.e.}, $\ell$ is such that ${\cal E} < R$, the resiliency capacity of the system is,
\begin{equation}
C_r^{BL}(\Gamma) = \sum_{i=b+1}^{k}(n-i)\beta,
\end{equation}
where $\beta = \Gamma/(n-1)$.
\end{theorem}

The condition ${\cal E} < R$ in  Theorem~\ref{thm:LtdCap} says that the eavesdropping capability of the adversary is insufficient to determine the message stored on
the DSS, {\em i.e.,} the adversary is not omniscient. This limitation in the adversary's knowledge enables every data collector to identify the erroneous symbols
introduced by the adversary and discard them, thus, resulting in erasures rather than errors. In this case also, identifying the erroneous symbols helps in the expurgation of
the system and  discarding the malicious nodes, as pointed out in Remark~\ref{rm:expurgation}.

The proof of Theorem~\ref{thm:LtdCap} is detailed in  Appendix~\ref{app:ltd} and is composed of two parts. In the first part, we assume that the hash table is
secure from the adversary and generalize the reasoning of Example~\ref{ex:Ltd_example} to show how the hash table can be used to identify, with high probability,
the erroneous symbols introduced by Charlie and thus decode correctly. In the second part, we demonstrate an efficient scheme to store the hash table securely and
reliably with a negligible sacrifice in the system capacity.



\section{Conclusion}\label{sec:conclusion}
In this paper we have considered the problem of securing a
distributed storage system under {\em repair dynamics} against eavesdropping and
 adversarial attacks. We proposed a new dynamical model
for the intrusion, wherein the adversary intrudes the system at
different time instances in order to exploit the system repair
dynamics to its own benefit. For the general model of an adversary
that can eavesdrop and/or maliciously change the data on some nodes
in the system, we investigate the problem of determining the {\em
secrecy capacity} and {\em resiliency capacity} of the system. We
provide upper bounds on the {\em
secrecy and resiliency} capacity and show their achievability in the
{\em bandwidth-limited regime}. General expressions of these capacities in addition to efficient decoding algorithms  remain an open problem.






\begin{appendix}
\subsection{Proof of Theorem~\ref{thm:UB_passive}}\label{app:UB_passive}

\begin{figure}[t]
\begin{center}
\resizebox{3in}{!} {\input{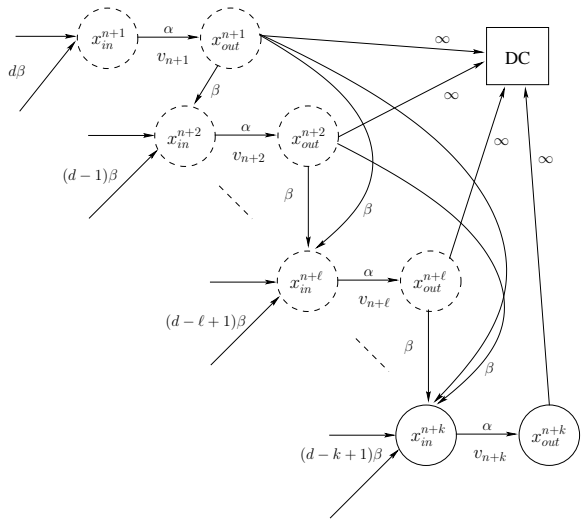}} \caption{ {\black
Part of the information flow graph corresponding to a DSS ${\cal D}(n,k,d)$, when
nodes $v_1,\dots,v_k$ fail successively and are replaced by nodes
$v_{n+1},\dots,v_{n+k}$. A data collector contacts these $k$ nodes and
wants to reconstruct the stored file. Nodes $v_{n+1},\dots,v_{n+\ell}$
shown with broken boundaries are compromised by Eve while they were
being repaired}.}\label{fig:converse}
\end{center}
\end{figure}

Consider a DSS $\mathcal{D}(n,k,d)$ with $\ell < k$, operating at point $(\alpha, \gamma)$ with $d\beta = \gamma$. Assume that
nodes $v_1, v_2,\dots,v_k$ have failed successively and were
replaced during the repair process by the nodes
$v_{n+1},v_{n+2},\dots,v_{n+k}$ respectively as shown in
the corresponding information flow graph ${\cal G}$ in Fig.~\ref{fig:converse}. Now suppose that Eve accesses the $\ell$ input nodes in the set $E=\{x^{n+1}_{in},x^{n+2}_{in},\hdots,x^{n+\ell}_{in}\} \subset V_{in}$ while they were being repaired. Consider also a data collector DC that downloads data from the $k$ output nodes in
$B=\{x^{n+1}_{out},x^{n+2}_{out},\hdots,x^{n+k}_{out}\} \in V^a_{out}$. The reconstruction property
of Eq.~\eqref{eq:reconstruction_condition} implies $H(S|C_B)=0$ and
the perfect secrecy condition in Eq.~\eqref{eq:secrecy_condition}
implies $H(S|D_E)=H(S)$. We can therefore write {\black
\begin{equation*}
\begin{split}
H(S) & = H(S|D_E)-H(S|C_B)\\
& \overset{(1)}\leq H(S|C_E)-H(S|C_B)\\
& \overset{(2)}= H(S|C_E) - H(S|C_E,C_{B\setminus E})\\
&= I(S,C_{B\setminus E}|C_E)\\
&\leq H(C_{B\setminus E}|C_E)\\
&= \sum_{i=\ell+1}^k H(C_{n+i}|C_{n+1},\dots,C_{n+i-1})\\
&\overset{(3)}\leq \sum_{i=\ell+1}^{k} \min\{(d-i+1)\beta,\alpha\}
\end{split}
\end{equation*}
} Inequality $(1)$ follows from  the Markov chain $S \rightarrow D_E \rightarrow C_E$ i.e., the stored data $C_E$ is dependent on $S$ only through the downloaded data
$D_E$, (2) from {\black $C_{B\setminus E}:=\{C_{n+\ell+1},\dots,C_{n+k}\}$,} (3) follows from the fact that each node can store at most $\alpha$ units, and for
each replacement node we have $H(C_i) \leq H(D_i) \leq d\beta$, also from the topology of the network (see Fig.~\ref{fig:converse}) where each node $x_{in}^{n+i}$
is connected to each of the nodes $x_{out}^{n+1},\dots,x_{out}^{n+i-1}$ by an edge of capacity $\beta$. The upper bound of Theorem~\ref{thm:UB_passive} then follows
directly from the definition of Eq.~\eqref{eq:secrecy_capacity}.


{
\subsection{Proof of Theorem~\ref{thm:cutset}}\label{app:omni_UB}
Consider a DSS $\mathcal{D}(n,k,d)$  operating at point $(\alpha, \gamma)$ with $d\beta = \gamma$, in the presence of an omniscient adversary that can control $b$ nodes, with $2b < k$. Assume that
nodes $v_{j+1},v_{j+2},\hdots,v_k$, for some $j$, $2b < j < k$, have failed consecutively and were replaced  by  nodes $v_{n+1},v_{n+2},\dots,v_{n+(k-j)}$, respectively. The information flow graph ${\cal G}$ of the  DSS corresponding to this sequence of  node failures and repairs  is shown in Fig.~\ref{fig:UBcut}. Consider a data collector
(Fig.~\ref{fig:UBcut}) that observes the stored data on the
$k$ nodes $v_{1},\dots,v_{j},v_{n+1},\hdots,v_{n+k-j}$. Consider also  the cut $C(V,\bar{V})$ with $\bar{V} = \{x^1_{out},\hdots,x^j_{out},x^{n+1}_{in},\hdots,x^{n+k-j}_{in}$, $x^{n+1}_{out},\hdots,x^{n+k-j}_{out}, \mbox{DC}\}$ that separates the source node $s$ from the data collector DC.
\begin{figure}[t]
\begin{center}
\resizebox{3in}{!} {\input{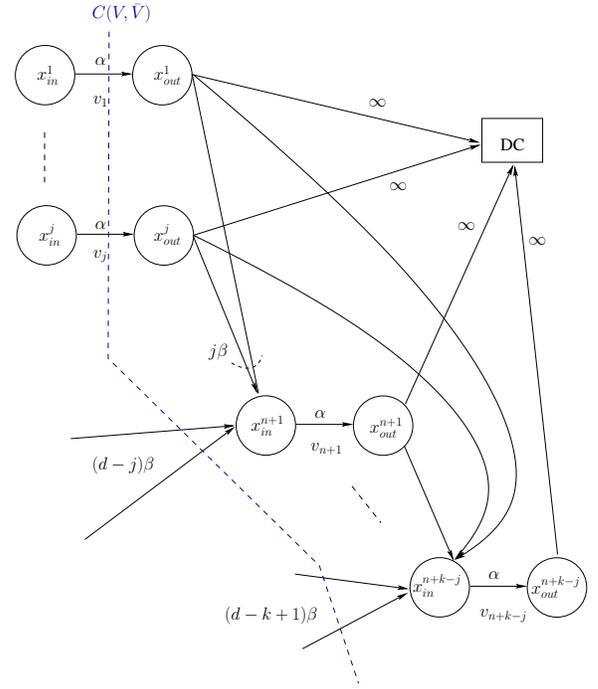}} \caption{Part of the
information flow graph corresponding to a DSS $(n,k,d)$ when nodes $v_{j+1},\hdots,v_k$ fail successively and are replaced by nodes $v_{n+1},\hdots,v_{n+k-j}$. A data collector connects to nodes $v_{1},\dots,v_{j},v_{n+1},\hdots,v_{n+k-j}$ to retrieve
the file.}\label{fig:UBcut}
\end{center}
\end{figure}
We group the edges belonging to this cut  into $3$ disjoint sets as follows:
\begin{enumerate}
\item  $E_1$: the set of edges outgoing from nodes $x^{p}_{in}, p =
1,\hdots,b$.
\item  $E_2$: the set of edges outgoing from nodes $x^{p}_{in}, p =
b+1,\hdots,2b$.
\item $E_3$:  the set of edges outgoing from nodes $x^{p}_{in}, p =
2b+1,\hdots,j$, in addition to the edges belonging  to the cut $C(V,\bar{V})$ that are  incoming to the nodes $x^{q}_{in}, q = n+1,\hdots,n+k-j$.
\end{enumerate}
Let $X_{E_i}(m), i = 1,2,3,$ be the symbols transmitted on the edges
in set $E_i$ corresponding to the stored message $m$. We claim that in the presence of an adversary controlling any $b$ nodes and for any two distinct messages $m_1 \neq m_2$ the following condition is necessary for the  DC to not make a decoding error:
\[
X_{E_3}(m_1) \neq X_{E_3}(m_2).
\]
Suppose that there exist two distinct messages $m_1 \neq m_2$
satisfying $X_{E_3}(m_1) = X_{E_3}(m_2)$. Now, if the symbols carried on the edges belonging to the cut $C(V,\bar{V})$ are $X_{E_1}(m_1), X_{E_2}(m_2)$ and $X_{E_3}(m_1) = X_{E_3}(m_2)$. Then, assuming all the messages  to be equally likely,  the
data collector will make a decoding error  with
probability at least $1/2$. This is true  since it will not be able to distinguish
between the following two cases:
\begin{itemize}
\item The true message is $m_2$ and the nodes $x^{1}_{in},\ \hdots,
\ x^{b}_{in}$ are controlled by the adversary Calvin who changed the transmitted symbols on the edges in the set $E_1$, from $X_{E_1}(m_2)$
to $X_{E_1}(m_1)$.
\item The true message is $m_1$ and the nodes $x^{b+1}_{in},\ \hdots,
\ x^{2b}_{in}$ are controlled by the adversary Calvin who changed the transmitted symbols on the edges in the set $E_2$, from $X_{E_2}(m_1)$
to $X_{E_2}(m_2)$.
\end{itemize}
Thus, the capacity of the DSS is upper bounded by the total capacity
of the edges in the set $E_3$, {\em i.e.},
\[
C_r(\alpha,\gamma) \leq \sum_{i=2b+1}^j \alpha + \sum_{i=j+1}^{k} (d-i+1)\beta, \ \ \ \ \ \ \ \ \  j=2b+1,\dots,k-1.
\]
The same analysis, as above, can be applied for $j=2b$ resulting in,
\[
C_r(\alpha,\gamma) \leq \sum_{i=2b+1}^{k} (d-i+1)\beta.
\]
And also for $j=k$, which gives,
\[
C_r(\alpha,\gamma) \leq \sum_{i=2b+1}^k \alpha.
\]
The bound in Theorem~\ref{thm:cutset} then follows by taking the minimum of all the above upper bounds obtained for $j=2b,\dots,k$. It can be easily seen that the
above argument extends to the case of $2b\geq k$ for which the set $E_3$ is empty and $C_r(\alpha,\gamma)=0.$ \hfill $\Box$ }

\subsection{Proof of Theorem~\ref{thm:LtdCap}} \label{app:ltd}

Consider a DSS $\mathcal{D}(n,k,d)$, with $d=n-1$, operating in the bandwidth-limited regime, in the presence of an adversary that can eavesdrop on $\ell$ nodes and control a subset of them of size $b$, $b\leq \ell$. As in the earlier proofs, we show the achievability for $\beta=1$, {\em i.e.}, $\Gamma = n-1$. Any larger values of $\beta$ or $\Gamma$ can be achieved by repeatedly applying the proposed scheme. Since there is no constraint on the node storage capacity $\alpha$ in bandwidth-limited regime, we choose $\alpha = n-1$. Let $\theta := \frac{n(n-1)}{2}$, $M:= \sum_{i=1}^k (n-i)$, $R:=\sum_{i=b+1}^k(n-i)$ and ${\cal E}:= \sum_{i=1}^{\ell}(n-i)$.

Our proof consists of two parts: 1) We assume that the hash table can be stored securely and reliably, and show an achievable scheme that can attain the resiliency capacity. 2) We present an efficient method to reliably
and securely store the hash table in the presence of a
limited-knowledge adversary Charlie.
\subsection*{C.1  Resiliency Capacity in the Limited-knowledge Case for the Bandwidth-Limited Regime}\label{app:achievable_ltd}
\subsubsection*{Code Construction} The code that we propose here is a generalization of the one used in Example~\ref{ex:Ltd_example} of Section~\ref{sec:limited_adversary}. It consists of an outer $(\theta,R)$ MDS code whose output $X=(\mbf{x_1},\hdots,\mbf{x_\theta})\in \mathbb{F}_{q^v}^\theta$   is  stored on the $n$ storage nodes using an inner RSKR-repetition code that enables exact repair in case of any node failure. As shown in Table~\ref{tab:Ltd_general_code}, each data packet $\mbf{x_i} \in \mathbb{F}_{q^v}, i=1,\dots,\theta$, is further appended with a hash vector $\mbf{h_i}=(h_{i,1},\dots,h_{i,\theta}) \in \mathbb{F}_q^{\theta}$. The values of these hashes are computed as follows,
\begin{eqnarray*}
{h}_{i,j} &=& \mbf{x_i}\mbf{x_j}^T,
\end{eqnarray*}
for $j = 1,2,\hdots,\theta$, where with abuse of notation $\mbf{x_i}$ also denotes the vector in $\mathbb{F}_q^{v}$ representing the corresponding element of
$\mathbb{F}_{q^v}$.  We assume for now that the hash values stored on the nodes are secure from Charlie who can neither observe nor corrupt them (as shown in the
next section). Although Charlie cannot directly observe the hash table, he can compute some of the hash values using the observed data packets on $\ell$
eavesdropped nodes and possibly introduce errors that are consistent with these hash values.

\begin{table}[h]
\begin{center}
\resizebox{3.5in}{!}{
\begin{tabular}{|c|c|c|c|c||c|c|c|c|}
  \hline
  Node & \multicolumn{4}{|c||} {data packet $\in \mathbb{F}_{q^v}$} & \multicolumn{4}{|c|}{hash $\in \mathbb{F}_q^{\theta}$}\\
  \hline
  $v_1$& $\mbf{x_1}$& $\mbf{x_2}$&$\hdots$& $\mbf{x_{n-1}}$             & $\mbf{h_1}$&$\mbf{h_2}$&$\hdots$&$\mbf{h_{n-1}}$\\
  \hline
  $v_2$& $\mbf{x_1}$& $\mbf{x_n}$&$\hdots$& $\mbf{x_{2n-3}}$            & $\mbf{h_1}$&$\mbf{h_n}$&$\hdots$&$\mbf{h_{2n-3}}$\\
  \hline
  $v_3$& $\mbf{x_2}$& $\mbf{x_n}$&$\hdots$& $\mbf{x_{3n-6}}$            & $\mbf{h_2}$&$\mbf{h_n}$&$\hdots$&$\mbf{h_{3n-6}}$\\
  \hline
  $\vdots$& $\hdots$& $\vdots$&$\ddots$& $\vdots$                       & $\vdots$&$\vdots$&$\ddots$&$\vdots$\\
  \hline
  $v_n$& $\mbf{x_{n-1}}$& $\mbf{x_{2n-3}}$&$\hdots$& $\mbf{x_{\theta}}$ & $\mbf{h_{n-1}}$&$\mbf{h_{2n-3}}$&$\hdots$&$\mbf{h_{\theta}}$\\
  \hline
\end{tabular}}
\end{center}
\caption{Schematic form of the code stored on the DSS $(n,k,d=n-1)$, along with the hash table that is not accessible to the adversary Charlie.}\label{tab:Ltd_general_code}
\end{table}

\subsubsection*{Decoding Logic}
A data collector accessing any $k$ nodes will observe a total of
$(n-1)k$ symbols and the corresponding hash vectors, where $\binom{k}{2}$ indices are repeated twice. As noted earlier, since the adversary can corrupt both of the stored symbols with same indices identically, the decoder focuses only on a set of $M = \sum_{i=1}^k (n-i)$ symbols with distinct indices along with their hash vectors
to make a decoding decision. These $M$ symbols form a codeword of an $(M,R)$ MDS code ${\cal X}$ possibly corrupted by errors introduced by the adversary.

Recall that Charlie can eavesdrop on a total of $\ell$ nodes and control some subset $b\leq\ell$ of these eavesdropped nodes in the system. Let $\mbf{y_i}, i = 1,\hdots,\theta$, denote the possibly corrupted version of the original data symbols $\mbf{x_i}$. We have $\mbf{y_i}=\mbf{x_i}+\mbf{e_i}$, where $\mbf{e_i}$ is the error introduced by Charlie on the symbols stored on the nodes he controls, and for rest of symbols $\mbf{e_i} = 0$. Without loss of generality, we suppose that the data collector observes nodes $v_1,\hdots,v_k$, {\em i.e.}, data
symbols $\mbf{y_i}$ and hash values $\mbf{h_i}, i \in \{1,2,\hdots,M\}$. The data collector observes the hash values with no errors since the hash table is assumed to be secure and reliable against the adversary. Let $H$ denote the observed ${M \times \theta}$ hash matrix having the vectors $\mbf{h_i} \in \mathbb{F}_q^{\theta}, i = 1,\hdots,M$ as rows. The data collector then computes its own $M \times M$ hash matrix $\hat{H}$ as
\[
\hat{H}_{ij}=\mbf{y_i}\mbf{y_j}^T, \ \ \ \ \ \ \ 1 \leq i,j \leq M
\]
from the observed $M$ data packets and compares it with the
corresponding entries in $H$. It generates an $M \times M$ comparison
table similar to Table~\ref{tab:checkmarks} in
Example~\ref{ex:Ltd_example}. In this table a ``$\checkmark$"
in  the $i$-th row and $j$-th column indicates that the computed hash and the observed
hash match, {\em i.e.}, $\hat{H}_{ij} = H_{ij}$, whereas ``$\times$"
indicates that $\hat{H}_{ij}\neq H_{ij}$ due to the errors introduced by the adversary.

The decoder then selects a set of $R$ symbols, among
$(\mbf{y_1},\dots,\mbf{y_M})$,  that index an $R\times
R$ sub-table of the comparison table with all its entries equal to
``$\checkmark$", and declares it as a {\em trusted set} with no errors. Then, it sets the rest of the $M-R$ observed symbols as
erased and proceeds to decode the obtained vector as a codeword of an  $(M,R)$ MDS code ${\cal X}$ with $M-R$ erasures. Since Charlie can control only $b$ nodes there always exists at least one set of size $M-\sum_{i=1}^{b}(n-i) = R$ symbols that generates a consistent hash sub-table of size $R\times R$ with ``$\checkmark$". Hence, the proposed decoder is guaranteed to stop. Next, we compute the probability that the above decoder decodes to an incorrect message.\\

\subsubsection*{Error Analysis} The proposed decoder may result in an error in decoding only if the chosen trusted
set of $R$ observed symbols contains at least one erroneous symbol,
say $\mbf{y_j} = \mbf{x_j} + \mbf{e_j}$, $\mbf{e_j} \neq 0$.
 Also, since $b \leq \ell$,
we have,
\begin{equation}\label{eq:b_control}
\sum_{i=1}^{b}(n-i) \leq \sum_{i=1}^{\ell}(n-i) < R,
\end{equation}
where the last inequality follows from our assumption (see Theorem~\ref{thm:LtdCap}) that the eavesdropping capability ${\cal E}$ is strictly less than the desired storage rate $R$. From equation~\eqref{eq:b_control}, it is clear that the
chosen trusted set contains at least one error-free symbol that is not observed by Charlie, say $\mbf{y_i} = \mbf{x_i} \notin E$. For this set to be a trusted set, it has to generate a consistent hash table of size $R \times R$.
In particular $\hat{H}_{ij} = H_{ij}$, {\em i.e.}, $\mbf{x_i}\mbf{e_j}^T =
0$.

Next, we compute the probability of such  event.  Let $E$ be the set of symbols in the codeword   $X$ that are observed by Charlie. Since $X$ is the output of a
$(\theta, R)$ MDS code and $|E| < R$,  any symbol $\mbf{x_i}$ of $X$ that does not belong to $ E$ is uniformly distributed in $\mathbb{F}_{q^v}$ conditioned on $E$,
{\em i.e.},
\begin{equation}\label{eq:cond_independence}
Pr(\mbf{x_i} = x_i|E) = \frac{1}{q^v} , \ \ \ \ \ x_i \in \mathbb{F}_{q^v}.
\end{equation}

Therefore, for any choice of $\mbf{e_j}$ that Charlie makes based on his observation $E$, there are $q^{v}$ equally likely choices of $\mbf{x_i}$ out of which
$q^{v-1}$ are orthogonal to the chosen $\mbf{e_j}$. Hence, the consistency condition of hash $\hat{H}_{i,j} = H_{i,j}$ is satisfied with probability,
\[
P_r(\mbf{x_i}\mbf{e_j}^T = 0|E,\mbf{e_j}) = \frac{1}{q},
\]
which goes to zero with increasing field size $q$.

 Note that if Charlie could observe the complete hash table, $\mbf{x_i}$ would  no more be independent of Charlie's observation. Then, as shown in  Example~\ref{ex:Ltd_example}, Charlie can always choose $\mbf{e_j}$ to belong to the orthogonal space of all possible choices of $\mbf{x_i}$, thus deceiving the proposed decoder. Therefore, it is crucial to keep the hash values secure from Charlie.

\subsubsection*{Rate Analysis} We encode $R$ information symbols in
$\mathbb{F}_{q^v}$ using a $(\theta,R)$ MDS code to form a codeword $(\mbf{x_1},\hdots,\mbf{x_{\theta}})$. For these symbols we construct a hash table of size
$\theta \times \theta$ with symbols in $\mathbb{F}_q$. Hence the total overhead of the hash table is $\frac{\theta^2}{R v} = {\cal O}(\frac{1}{v})$ per information
symbol which goes to zero with increasing block length $v$. Hence, asymptotically in block length $v$, these codes achieve the capacity of Theorem~\ref{thm:LtdCap}.


\subsection*{C.2  Reliable and Secure Storage of the Hash Table}

\label{app:hash_security} The scheme described here for storing the hash table securely and reliably is along the parallel lines of the scheme proposed \cite{YSJL10}\footnote{
The scheme of \cite{YSJL10} is matrix-based and is designed for networks where intermediate nodes perform random network coding. Our scheme here can be regarded as a simple vector version of the one  in \cite{YSJL10}. This simplification is possible due to the special structure of the networks (information flow graphs) representing distributed storage systems  in conjunction with the RSKR-repetition codes that limit coding in these networks to the source.} in the context of securing multicast networks. It aims at storing $1$ bit of
information securely and reliably. The scheme can then be repeated to store the complete hash table which, as shown in the previous section, is of constant size and independent of the block length $v$ of the information symbols. The total overhead incurred by this scheme can be then  made arbitrarily small by increasing $v$.\\

\subsubsection*{Code Construction} Let $G = \left(\begin{array}{c}
G_K \\
G_S \\
\end{array}
\right)$ be a generator matrix of a $(\theta,M)$ nested MDS code over the
finite field $\mathbb{F}_q$ (symbols in the hash table also belong to the same field). The matrix $G_K$ in itself is a generator matrix of a $(\theta, {\cal E})$ MDS code over $\mathbb{F}_q$. If the bit to be stored is ``$1$'' then choose a vector $S$ randomly and uniformly
from $\mathbb{F}_q^{M-{\cal E}}$, otherwise, set $S=0 \in
\mathbb{F}_q^{M-{\cal E}}$. Let ${\cal K} = (K_1\hdots,K_{\cal E})$
denote ${\cal E}$ random keys mutually independent and each uniformly distributed over $\mathbb{F}_q$. Now, we form the vector $X \in
\mathbb{F}_q^{\theta}$ to be stored on the DSS as part of the hash table by   ``mixing''  $S$
with the random keys using the nested MDS code as,
\begin{eqnarray*}
X &=& {\cal K} G_K + S G_S.
\end{eqnarray*}

This encoded vector $X \in \mathbb{F}_q^{\theta}$ is then stored on the $(n,k,d)$ DSS using the RSKR-repetition code as shown in
Fig.~\ref{fig:Gamma_repetition}. The RSKR-repetition structure  allows the exact repair of a node in case of failure as explained in Section~\ref{sec:Passive_adversary}.\\

\subsubsection*{Security Analysis}\label{sec:security_analysis} The coding scheme used here is same as the one in
Section~\ref{sec:passive_achievability} that discusses passive
adversary and hence the vector $S$, which is of the appropriate rate $M-{\cal E}$, is perfectly secure from Charlie
eavesdropping on $\ell$ nodes. The perfect secrecy of $S$ implies the perfect secrecy of the hash bit.

Next we describe a decoding algorithm that the data collector uses to
decode the stored bit with high probability of success even in the
presence of errors introduced by Charlie controlling $b$ nodes.\\

    %

\subsubsection*{Decoding Logic} We denote by ${\mathbb D}$  the decoder used by the data collector to recover the stored bit belonging to the hash table. ${\mathbb D}$ implements the same decoding steps as the decoder of Section~\ref{sec:omni_decoding}, of omniscient adversary, except for the decision rule that determines the output.
The input to ${\mathbb D}$ is the data observed by the data collector accessing $k$ nodes which is formed of $k\alpha = k(n-1)$ symbols, among which $\binom{k}{2}$ pairs have the same indices. The decoder  executes the following steps:

\begin{enumerate}
    \item ${\mathbb D}$ selects any set of $M$ symbols having  distinct
    indices  among the observed $k\alpha$   symbols. These symbols are grouped   in a vector  $Y \in \mathbb{F}_q^M$ which can be written as
    \begin{eqnarray*}
    Y &=& {\cal K} \bar{G}_K  + S \bar{G}_S+\mbf{e},
    \end{eqnarray*}
    where $\bar{G}_K $ and $\bar{G}_S$ are submatrices of $G_K$ and $G_S$ of size ${\cal E}\times M$ and $(M-{\cal E})\times M$, respectively. The vector   $\mbf{e} \in \mathbb{F}_q^{M}$, with up to $\sum_{i=1}^{b}(n-i)$ non-zero terms, is the error vector that accounts for the errors introduced by the adversary.
\item Let $B, |B|=b$, denote the set of storage nodes
controlled by the adversary. Again, due to the exact repair property of the RSKR-repetition code it is sufficient to consider $B\subset \{v_1,\dots,v_n\}$ with $|B|=b$. For each such set
$B$, let $I_B \subset \{1,2,\hdots,\theta\}$ denote the set of
indices of the symbols stored on the nodes in $B$.
\item For each possible $B \subset \{v_1,v_2,\hdots,v_n\}$, $|B|=b$, $\mathbb{D}$ punctures $Y$ with pattern $I_B$ to obtain $Y_{I_B}$ as
    \begin{eqnarray*}
    Y_{I_B} &=& {\cal K} \bar{G}_{K_{I_B}}  + S \bar{G}_{S_{I_B}}+\mbf{e}_{I_B},
\end{eqnarray*}
where $\bar{G}_{K_{I_B}}$ and $\bar{G}_{S_{I_B}}$ are the submatrices of $\bar{G}_K $ and $\bar{G}_S$ obtained by deleting the columns corresponding to the punctured elements of $Y$, and $\mbf{e}_{I_B}$ is the punctured error vector.
    \item  ${\mathbb D}$ checks whether $Y_{I_B}$ is a valid codeword of the code generated by the matrix $\bar{G}_{K_{I_B}}$ by checking whether the corresponding syndrome is zero.
    \item The decoder $\mathbb{D}$ repeats steps $3)$ and $4)$ for each of the $\binom{n}{b}$ sets $B$ until  the syndrome obtained in step $4)$ is zero. In this case,   ${\mathbb D}$ declares that bit ``0'' was stored. Otherwise, if for all possible values of $B$ no zero syndrome is obtained,  ${\mathbb D}$ declares that ``1'' was stored.\\
\end{enumerate}

\subsubsection*{Error Analysis} We do the error analysis of the above decoding logic considering two different cases
based on the value of the stored hash bit.

\begin{itemize}
%

\item {{\em Hash bit `0':}} We will show that when the stored information bit  is `0', the decoder ${\mathbb D}$ makes no error. In fact, this case corresponds to $S=0$ and, thus, $Y = {\cal K} \bar{G}_K+\mbf{e}$. Let $B^*$ be the actual set of nodes controlled by Charlie. Then, there is at least one set $B=B^*$ for which $Y_{I_B^*} = {\cal K} \bar{G}_{K_{I_B^*}}$, since $\mbf{e}_{I_{B^*}} = 0$. As a result, the decoder always outputs ``0''.

\item {{\em Hash bit `1':}} Information bit `1' corresponds to
\begin{eqnarray*}
Y &=& {\cal K} \bar{G}_K  + S \bar{G}_S + \mbf{e},
\end{eqnarray*}
where $({\cal K}, S)$ is a uniformly random vector in $\mathbb{F}_q^{M}$ and $\mbf{e} \in \mathbb{F}_q^M$ is the error
vector introduced by Charlie. Note that the matrix $G$ is a generator matrix of a $(\theta,M)$ MDS code, hence the  $M \times M$ sub-matrix
$\bar{G} := \left(\begin{array}{c}
                  \bar{G}_K \\
                  \bar{G}_S
                \end{array}\right)$ is invertible. Thus, we can write
\begin{eqnarray} \label{eq:error_event}
Y &=& ({\cal K} + \mbf{e}_K)\bar{G}_K  + (S + \mbf{e}_S) \bar{G}_S,
\end{eqnarray}
where $\mbf{e}_K,\mbf{e}_S$ are the coefficients of the error vector
$\mbf{e}$ in terms of the basis corresponding to the rows of
$\bar{G}_K,\bar{G}_S$. We have already shown in the security analysis above, that $S$ is perfectly secure from Charlie's observation. Hence $S + \mbf{e}_S$ is a uniformly random vector in $\mathbb{F}_q^{M-{\cal E}}$.

Consider any set $B \subset \{v_1,\hdots,v_n\}$ of cardinality $|B|=b$ with index set $I_B$. Then, $|I_B| = \sum_{i=1}^b (n-i)$, hence the matrix $\bar{G}_{I_B}$ obtained by deleting the columns of $\bar{G}$ corresponding to the indices $I_B$ has $R = M-|I_B|$ or more columns. Now, the matrix $\bar{G}_{K_{I_B}}$ is a generator of an  $(M,{\cal E})$ MDS code and ${\cal E} < R$ (Theorem~\ref{thm:LtdCap}). Hence, the rank of $\bar{G}_{K_{I_B}}$ is ${\cal E}$. This,  along with the fact that $\bar{G}$ is an invertible matrix, implies that the rank of matrix $\bar{G}_{S_{I_B}}$ is $R - {\cal E}$ or more. The probability, that the syndrome computed in the step $4)$ of the proposed decoding logic for this set $B$ is equal to zero, is equal to the  probability of the event that a uniformly random vector $(S + \mbf{e}_{S})$ lies in the space orthogonal to the span of columns of $\bar{G}_{S_{I_B}}$. This probability is upper bounded by $1/q^{R-{\cal E}}$.

Now applying the  union bound to all $\binom{n}{b}$ choices of the set $B$ that the decoder attempts, the probability of error can be upper
bounded by,
\[
\lim_{q\rightarrow\infty} \frac{\binom{n}{b}}{q^{R-{\cal E}}} \rightarrow 0
\]
which goes to zero with increasing the field size $q$.\\
\end{itemize}

\subsubsection*{Rate Analysis} In the code proposed above to store the hash values securely and reliably we need ${\theta}$ symbols in $\mathbb{F}_q$ for each $1$ bit of hash information. Also, in the previous section we showed that the total size of the hash table of interest is $\theta^2$ symbols in $\mathbb{F}_q$. Thus, the total overhead of the proposed code to store the hash table is $\theta^3\log{q}$ symbols of $\mathbb{F}_q$, that is independent of the block length $v$ of information packets.

Thus, we have shown how the hash table described in Table~\ref{tab:Ltd_general_code} can be
stored on the DSS with a negligible overhead and is guaranteed with a high probability to  be secret and resilient to the adversary provided that
 field size $q$ and block length $v$ are large enough.


\begin{table*}[t]
\begin{center}
\begin{tabular}{|c|l|}
  \hline
  Notation  & Explanation\\
  \hline
  \hline
  {$\cal G$} & Information flow graph of a distributed storage system.\\
  \hline
  {$\cal V$} & {Set of nodes in the information flow graph.} \\
  \hline
  {$C(V,\bar{V})$} & {Cut partitioning the set of nodes ${\cal V}$ in a graph into two sets $V \subset {\cal V}$ and ${\bar V} = {\cal V} \setminus V$.} \\
  \hline
  {$S$} & {Random variable representing an incompressible source file.} \\
  \hline

  {$n$} & {Total number of active nodes in a distributed storage system.} \\
  \hline
  {$k$} & {Number of nodes a data collector connects to in order to retrieve the source file.} \\
  \hline
  {$d$} & {Number of nodes a new replacement node connects to during the repair process.} \\
  \hline
  {$\alpha$} & {Storage capacity at each storage node in a distributed storage system.} \\
  \hline
  {$\beta$} & {Amount of data downloaded from every node participating in the repair process.} \\
  \hline
  {$\gamma$} & {The total amount of data downloaded during the repair process i.e., repair bandwidth.} \\
  \hline
  {$\Gamma$} & {Upper limit on the repair bandwidth in the bandwidth-limited regime.} \\
  \hline

  {$D_i$} & {All the data$\setminus$messages downloaded on the replacement node $v_i$ during the repair process.} \\
  \hline
  {$C_i$} & {Data stored on the node $v_i$.} \\
  \hline

  {$R$} & {Desired or achieved storage rate.} \\
  \hline
  {$M$} & {Capacity of the distributed storage system in the absence of an adversary.} \\
  \hline
  {$\mbf{x_i}$} & {Data symbol or packet stored on a distributed storage system.} \\
  \hline
  {$\mbf{y_i}$} & {Data symbol or packet, possibly corrupted by an adversary, observed by a data collector.} \\
  \hline

  {$\ell$} & {Number of nodes an adversary can eavesdrop on in a distributed storage system.} \\
  \hline
  {$b$} & {Number of nodes an active adversary can maliciously control.} \\
  \hline
  {$E$} & {A set of symbols$\setminus$nodes observed by an adversary by eavesdropping on $\ell$ nodes.} \\
  \hline
  {$C_s$} & {Secrecy capacity of a distributed storage system.} \\
  \hline
  {$C_r$} & {Resiliency capacity of a distributed storage system.} \\
  \hline
\end{tabular}\caption{Table of Important Notations}
\end{center}
\end{table*}
\end{appendix}




\bibliographystyle{ieeetr}

\begin{IEEEbiographynophoto}{Sameer Pawar} received the M.S. degree in
electrical engineering from Indian Institute of Science (IISc),
Bangalore, India, in 2005. Since 2007, he has been with the
Department of Electrical Engineering and Computer Science in the
University of California at Berkeley. Prior to that, he had been
with the Communications Department, Infineon Technologies India. His
research interests include information theory and Coding theory for
Storage and communication systems. He is recipient of Gold Medal for
the Best Masters thesis in Electrical Division in IISc.
\end{IEEEbiographynophoto}

\begin{IEEEbiographynophoto}{Salim El Rouayheb}(S'07–M'09) received the Diploma degree in electrical engineering from the Lebanese University, Faculty of Engineering, Roumieh, Lebanon, in 2002, and the M.S. degree in computer and communications engineering from the American University of Beirut, Lebanon, in 2004. He received the Ph.D. degree in electrical engineering from Texas A\&M University, College Station, in 2009.
He is currently a Postdoctoral Research Fellow with the Electrical Engineering and Computer Science Department, University of California, Berkeley. His research interests lie in the broad area of communications with a focus on reliable an secure distributed information systems and on the algorithmic and information-theoretic aspects of networking.
\end{IEEEbiographynophoto}

\begin{IEEEbiographynophoto}{Kannan Ramchandran} is a Professor of Electrical Engineering and Computer
Science at the University of California at Berkeley, where he has been since
1999. Prior to that, he was with the University of Illinois at Urbana-Champaign
from 1993 to 1999, and was at AT\&T Bell Laboratories from 1984 to 1990. His
current research interests include distributed signal processing algorithms for
wireless sensor and ad hoc networks, multimedia and peer-to-peer networking,
multi-user information and communication theory, and wavelets and multi-resolution
signal and image processing. Prof. Ramchandran is a Fellow of the IEEE.
His research awards include the Elaihu Jury award for the best doctoral thesis
in the systems area at Columbia University, the NSF CAREER award, the ONR
and ARO Young Investigator Awards, two Best Paper awards from the IEEE
Signal Processing Society, a Hank Magnuski Scholar award for excellence in
junior faculty at the University of Illinois, and an Okawa Foundation Prize for
excellence in research at Berkeley. He is a Fellow of the IEEE. He has published
extensively in his field, holds 8 patents, serves as an active consultant to industry,
and has held various editorial and Technical Program Committee positions.
\end{IEEEbiographynophoto}

\end{document}